\documentclass{iopart}

\usepackage{graphicx}

\def\la{\hbox{{\lower -2.5pt\hbox{$<$}}\hskip -8pt\raise
-2.5pt\hbox{$\sim$}}}
\def\ga{\hbox{{\lower -2.5pt\hbox{$>$}}\hskip -8pt\raise
-2.5pt\hbox{$\sim$}}}

\begin{document}

\begin{center}\title{Cosmogenic Neutrinos from the propagation of Ultra High Energy Nuclei}\end{center}

\author{D. Allard$^{1,2}$, M. Ave$^{1,4}$, N. Busca$^{1,3}$, M. A. Malkan$^{5}$, A. V.\ Olinto$^{1,2,4}$, E. Parizot$^6$, F. W. Stecker$^{5,7}$, and T. Yamamoto $^{1,4}$}
\address{$^1$Kavli Institute of Cosmological Physics}
\address{$^2$Department of Astronomy and Astrophysics}
\address{$^3$Department of Physics}
\address{$^4$Enrico Fermi Institute, \\
The University of Chicago, 5640 S. Ellis, Chicago, IL60637, USA}
\address{$^5$University of California, Los Angeles, CA 90095-1547, USA}
\address{$^6$Institut de Physique Nucl\'eaire d'Orsay, IN2P3-CNRS\\
Universit\'e Paris-Sud, 91406 Orsay Cedex, France}
\address{$^7$NASA/Goddard Space Flight Center\\
 Department of Physics and Astronomy, University of California, LosAngeles}

 \date{\today}

\begin{abstract}
We calculate the flux of  neutrinos generated by the propagation of ultra-high energy
nuclei over cosmological distances. The propagation takes into account the interactions with cosmic background radiations including the CMB and the most recent estimates of higher energy (infra-red, optical, and ultra violet) backgrounds. We assume that the composition of ultra-high energy cosmic rays (UHECRs) at the source is the same as the observed one at low
energies. This assumption fits well the present data at the highest energies. 
We compare the cosmogenic neutrino flux from mixed composition sources to pure proton sources. 
We find that the neutrino flux in the mixed composition case has a high energy peak, mainly due to photopion production off CMB photons, of similar shape and amplitude to the proton case. At low energies both composition cases have significant neutrino flux with a peak around $10^{14.5}$ eV  due to the higher energy backgrounds. The mixed composition case induces a higher flux of neutrinos at energies below $10^{13}$ eV due to the neutron decay component that extends down to  low energies. Detection of diffuse neutrino fluxes at ultra high energies can strongly constrain the source distribution of UHECR whereas fluxes at lower energies could be used to constrain confinement of VHE and UHE cosmic rays if combined with composition analysis from cosmic ray experiments.
\end{abstract}

\maketitle

\section{Introduction}

Mass composition measurements are key to solving the mystery of the origin of ultra-high energy cosmic rays (UHECRs). If UHECRs originate in extragalactic astrophysical accelerators, the observed composition of UHECRs should be primarily protons and light nuclei (see, e.g., \cite{Allard05a}), while heavier nuclei could indicate a Galactic origin (see, e.g., \cite{BEO00}), and photon primaries could indicate top-down scenarios (see \cite{BhatSigl00} for a review). Future composition measurements  should be able to differentiate between alternative scenarios for the origin of UHECRs and to determine the transition between Galactic and extragalactic cosmic rays. 

Present studies of cosmic ray composition indicate a dominance of heavier nuclei (from $^4$He to $^{56}$Fe) around the knee region \cite{Kascade} ($\sim 10^{16} - 10^{17}$ eV) followed by a tendency toward lighter nuclei around  $\sim 10^{18}$ eV and above \cite{FlysEye,HiRes}.
However, the composition of cosmic rays is notoriously difficult to determine at high energies. A number of airshower observables are composition dependent, such as the fraction of muons in the shower,  the shower maximum, and the fluctuations of shower maximum. These observables require observatories with large statistics, control of systematics, and a large dynamic range in observed energies. In addition,  comparisons of observations with model predictions depend on simulations of hadronic interactions in an energy and rapidity range that have not been tested by accelerator experiments.  Current efforts in  improving hadronic interaction models and the construction of the Pierre Auger Observatory \cite{PAO}  should lead to a much better understanding of UHECR composition. 

Another key observation that should become feasible in the next few years is the detection of cosmic neutrinos at high energies. UHECRs are of great interest as a source of high energy neutrinos that are almost guaranteed by their interactions  with the cosmic background radiation. Cosmic rays at the highest energies produce pions off the cosmic microwave background (CMB) giving rise to the Greisen-Zatsepin-Kuzmin feature in the cosmic ray spectrum \cite{GZK66}. The decay of the charged pions produced by the  interactions  with the CMB generates neutrinos \cite{BereOriginal,FWS,cosmogenic_nu}  that are often called cosmogenic, GZK, or photopion neutrinos.    Furthermore, it has been recently shown \cite{DDMSS05}, that the interactions of protons with the infra-red and optical intergalactic backgrounds  produce a large amount of cosmogenic neutrinos at lower energies. 

Most predictions of the cosmogenic neutrino flux assume that the UHECR primaries are protons. More recently cosmogenic neutrino fluxes were calculated for other primaries such as pure $^{56}$Fe \cite{Ave05a,Ave05b,THS05}, $^4$He, and $^{16}$O \cite{THS05}. Here we calculate the cosmogenic neutrino flux for sources that inject a mixture of primaries with the same initial abundances as the observed Galactic cosmic rays and compare with the pure proton case. The mixed composition model was proposed in  \cite{Allard05a} and studied in detail in \cite{Allard05b}, where a comparison between the predicted spectrum and composition for this model is contrasted with the pure proton assumption.  We briefly review the mixed composition model  and describe the adopted source evolution models in \S2. In order to calculate the neutrino flux, we propagate the injected primaries and their daughter nuclei  from cosmological distances to Earth calculating their interactions with cosmic backgrounds (\S 3). We simultaneously calculate  the generated flux of neutrinos for composition at injection (pure proton or mixed) and different source evolution scenarios  (\S4).   The generated flux of neutrinos for both composition hypotheses is normalized by comparing the predicted UHECR spectrum with present observations.  We discuss the implications of our results for ongoing and future experiments in \S 5 and conclude in \S6.

\section{Mixed Composition UHECR and Source Evolution Models}

In order to compare the predicted cosmogenic neutrino flux for a mixed composition scenario with previous models, we study the neutrino production for two UHECR injection compositions:  pure proton and mixed composition. The mixed nuclei case assumes that the injection composition at the source matches the abundances observed in Galactic cosmic rays at lower energies as in 
\cite{DuVernois}.
 The source is assumed to emit a power law spectrum with spectral index $\alpha$ such that the number of nuclei $i$ emitted in the energy range $ [E,E+dE]$ is:
\begin{equation}
n_{i}(E)=x_{i}A_{i}^{\alpha-1}\kappa E^{-\alpha}dE \ ,
\label{eq:2}
\end{equation}
where A is the mass number of a given nucleus,  $\kappa$ is a normalization constant, and $x_{i}$ is the abundance of  species $i$ (given in \cite{DuVernois}).

\begin{figure*}[ht]\centering\hfill~\hfill\includegraphics[width=6cm]{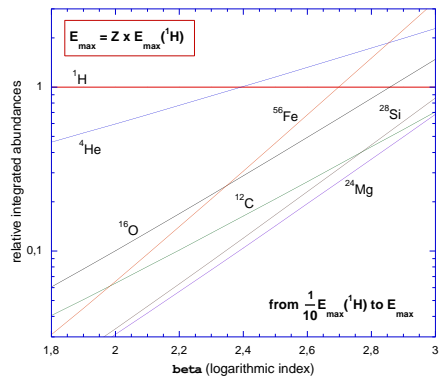}
\hfill~\caption{Evolution of the integrated relative abundances at the source normalized to protons as a function of the injection spectral index between $E_{max}/10$ and $E_{max}$.  }
\label{Abond}\end{figure*}

In the mixed composition model of \cite{Allard05a} , the maximum  energy at injection of each species is set to
\begin{equation}
E_{max,i}=Z_{i}  E_{max}(^{1}H)  \ ,
\label{eq:4}
\end{equation}
i.e.,  the maximun energy is proportional to the charge number $Z$ of a given nucleus. This assumption is reasonable if the maximun energy at the source is limited by the confinment of particles. 
The evolution of the relative abundances under these assumptions as a function of the injection spectral index is displayed in Fig.~\ref{Abond}. 
For the neutrino flux calculation, we first set the maximum proton energy $E_{max}(^{1}H) = 10^{20.5}$ eV for the pure proton case and use the same input $E_{max}(^{1}H)$ in eq. \ref{eq:4} to set the mixed composition case,  which gives a maximum energy for iron of $\sim 10^{22}$  eV.  We assume an exponential cut-off at the source above $E_{max}(Z)$. We discuss the effect of changes in $E_{max}$ to the expected neutrino flux below.

The predicted neutrino flux is strongly dependent on the choice of source redshift evolution. We consider four source evolution models in our predictions.  First, we assume no evolution with redshift (hereafter called uniform distribution). Second, we consider a  redshift dependence proportional to the old estimate of the star formation rate (SFR) of \cite{Madau96} (hereafter oSFR), which evolves as $(1+z)^n$ for $ z < 1.9$, followed by a constant $(2.9)^n$ between $1.9 < z < 2.7$, and an exponential cutoff $(2.9)^n exp(1 - z/2.7)$ for $z > 2.7$ and with $n=3$ \cite{Waxman} as assumed by most previous cosmogenic neutrino flux calculations. Our third assumption is based on a more recent estimate of the SFR evolution (hereafter new SFR or nSFR) that can be deduced from \cite{Bunker04} and evolves as $(1+z)^3$ for $ z < 1.3$, followed by a constant rate between $1.3 < z < 6$, followed by a sharp cut-off. Finally, we use a stronger source evolution (hereafter called strong evolution) favored by the recent infra-red survey of the Spitzer telescope \cite{Perez05}. We use the following parametrization  of their model: $(1+z)^4$ for $ z < 1$, followed by a constant rate between $1 < z < 6$, followed by a sharp cut-off.  The two latter source evolution hypothesis were recently used in \cite{DDMSS05}.

\section{Interactions of protons and nuclei with photon backgrounds and neutrino emission}

\begin{figure*}[ht]\centering\hfill~\hfill\includegraphics[width=6cm]{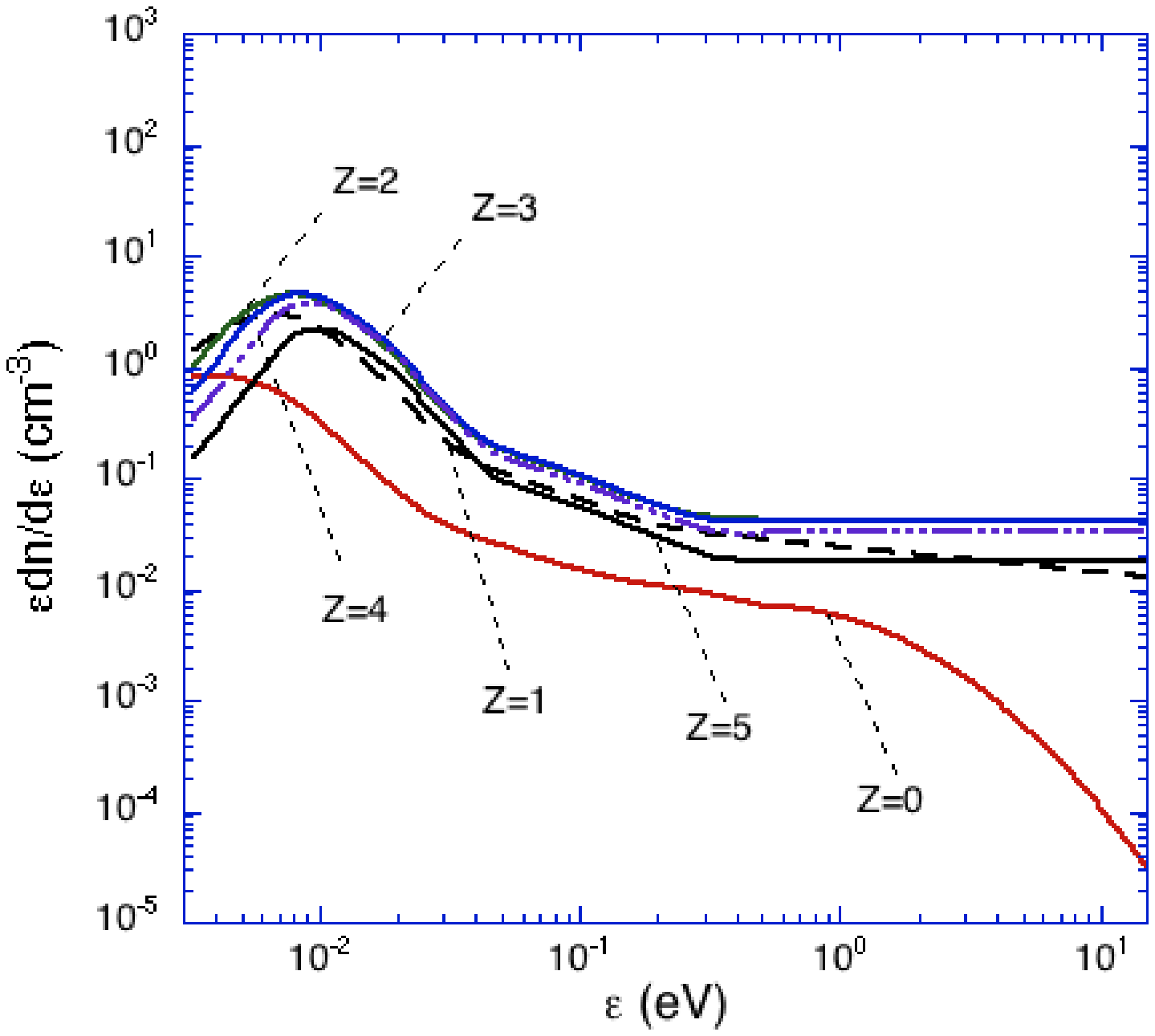}
\hfill\includegraphics[width=6cm]{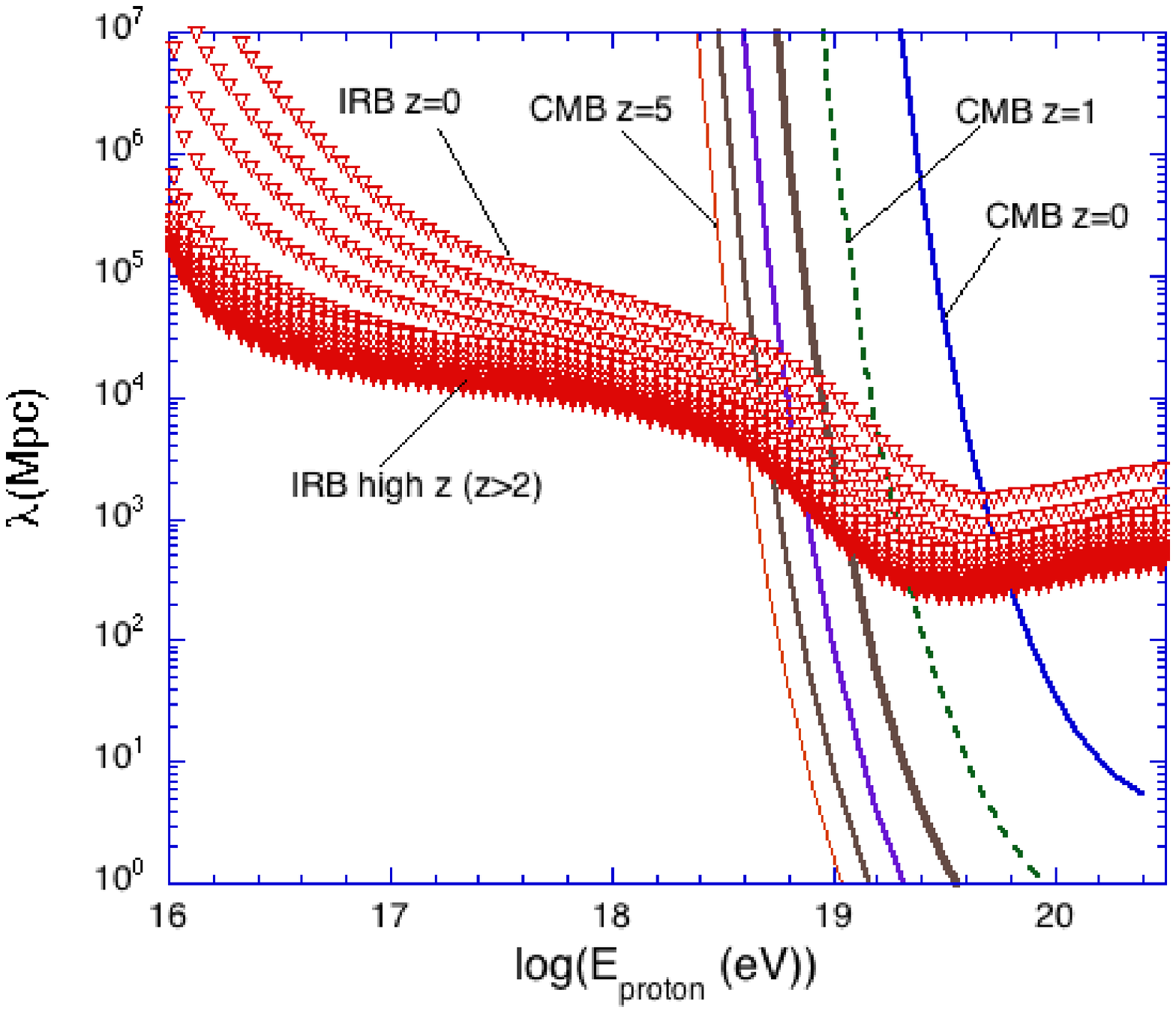}\hfill~\caption{Left: Evolution of the differential density of the IR/Opt/UV background for redshifts between 0 and 5 \protect\cite{MS05}. Right: Evolution of the proton mean free path for photo-pion production with the IR/Opt/UV and the CMB at redshifts between 0 and 5}\label{background}\end{figure*}

In the following sections we consider the interaction of protons and nuclei with the CMB and the infra-red, optical and ultraviolet backgrounds (hereafter we group these three backgrounds under IR/Opt/UV for short). The effect of IR/Opt/UV photons on the propagated UHECR  spectrum of pure  proton sources is negligible, however, as shown in \cite{DDMSS05},  these additional backgrounds have a significant effect on the neutrino flux associated with UHECRs.
To model the IR/Opt/UV backgrounds and their cosmological evolution, we use the latest estimate of \cite{MS05} which is based the earlier work in Refs.
\cite{MS98} and \cite{MS01} updated with  recent data on history of the star formation rate and the evolution of galaxy luminosity functions.
We use IR/Opt/UV calculated at 26 different redshifts ($\Delta z=0.2$) between 0 and 5. 
The differential density of the IR/Opt/UV is shown on Fig.~\ref{background}a. In general, the IR/Opt/UV has a much milder cosmological evolution when compared to the CMB. However, the evolution of the photon background in the UV range is notable for redshifts between 1 and 0. The UV background gets fainter at lower redshifts due to the aging of the stellar population which results in the death of the stars with the shortest lifetimes which produce all of the UV emission \cite{SS98}.The basic reason is that most galaxies were forming stars at much higher rates at z=1 than they are today (e.g., \cite{MS01}). The UV background has a sharp cut-off above 13.6 eV due to the Lyman limit absorption from HI observed in galaxies \cite{MKW}. In the following, we assume this edge to be black, i.e., we assume that there are no photons above 13.6 eV.  

Protons and nuclei propagating in the extragalactic medium interact with CMB and  IR/Opt/UV background  photons. These interactions produce features in the propagated UHECR spectrum such as the GZK cutoff \cite{GZK66} and their decay products generate the cosmogenic neutrino flux. The proton mean free paths for photo-pion interactions with the CMB and IR/Opt/UV background at different redshifts are displayed on Fig.~\ref{background}b. The evolution of the mean free path with redshift is  strong in the case of the CMB, which implies that the dominant background strongly depends on the source evolution assumed. It is also important to note that the photo-pion production off IR/Opt/UV photons  competes with the pair production process off CMB photons which is also evolving with redshift. Due to the slow evolution of the density of the IR/Opt/UV background with the energy, the mean free path above the interaction threshold (determined by the UV break of the background) evolves slowly. A sharper decrease of the mean free path is visible at higher energies when protons start to interaction with photons of the far IR bump. At very high energies (above $\sim 5  \times 10^{19} (1+z) $ eV) the CMB contribution starts to dominate the mean free path evolution and the effect of the IR/Opt/UV photons becomes  negligible. 

\begin{figure*}[ht]\centering\hfill~\hfill\includegraphics[width=6cm]{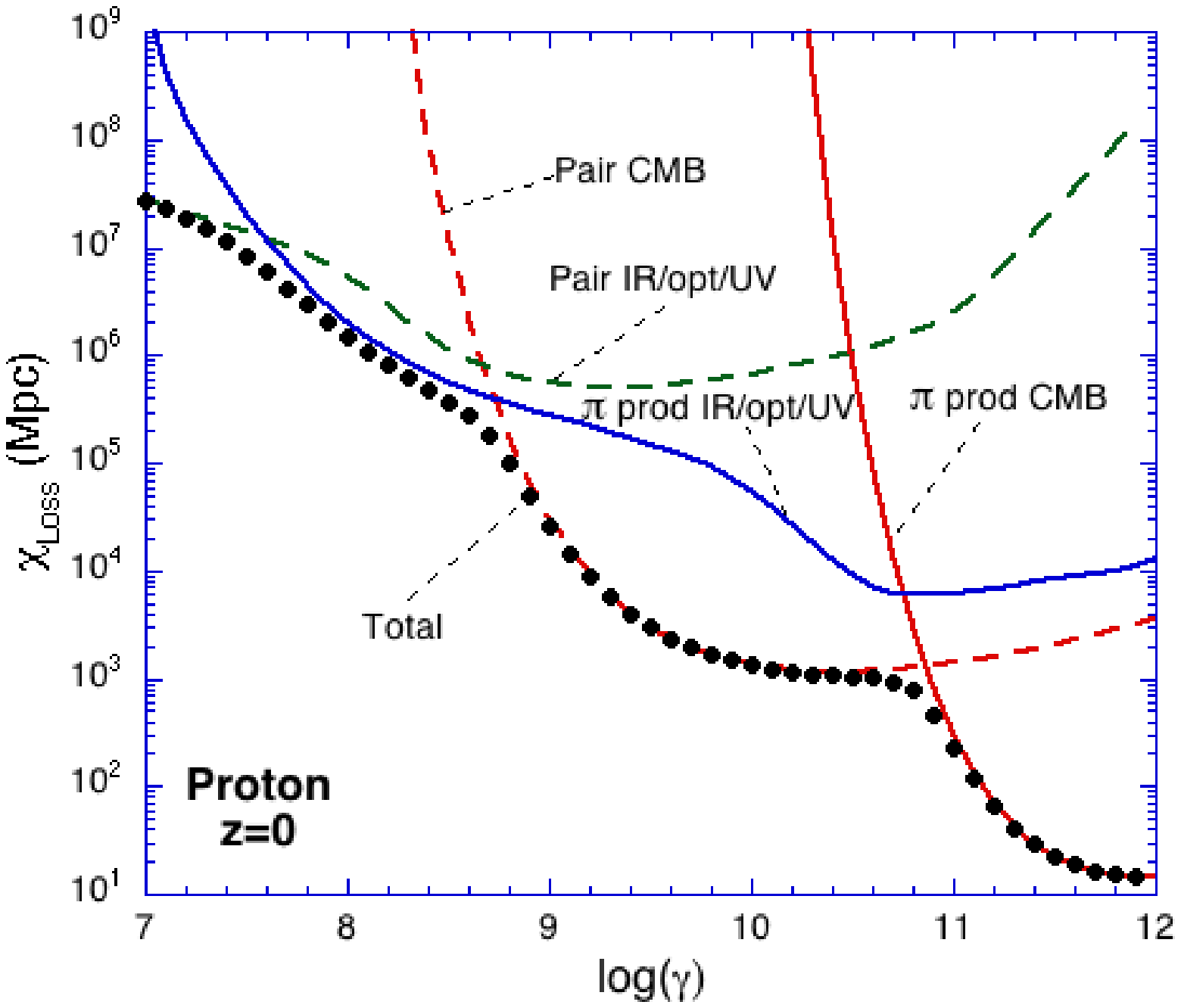}
\hfill\includegraphics[width=6cm]{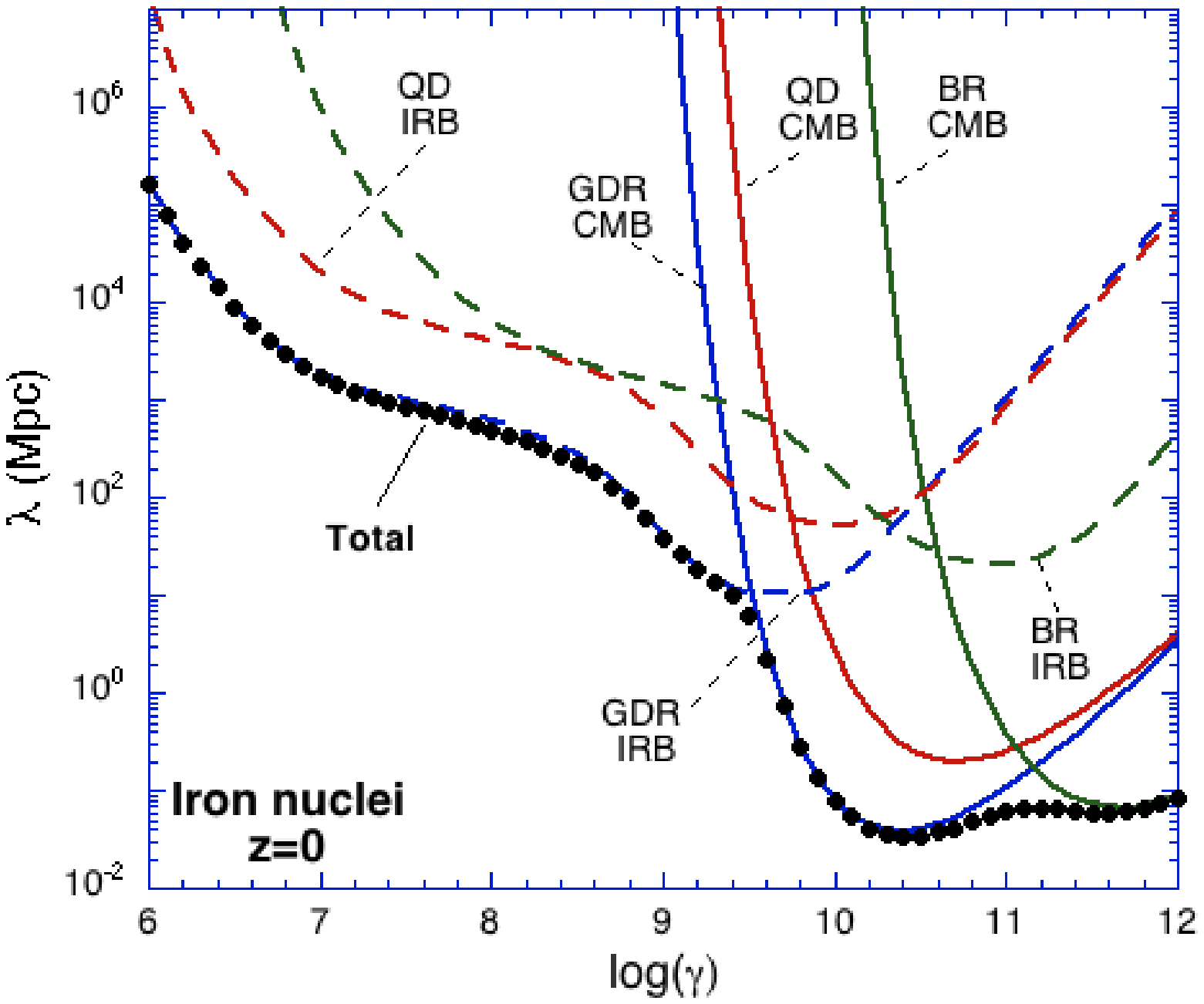}\hfill~\caption{Left: Evolution of the attenuation length of protons as a function of the energy at $z=0$. The contribution of pair production and pion production off the CMB and IR/Opt/UV  are separated. Right: Evolution of the iron nucleus mean free path for the different photoerosion processes and interactions with the CMB and IR/Opt/UV photons at $z=0$}\label{MFP1}\end{figure*}

Fig.~\ref{MFP1}a shows the contribution of the different processes with the different backgrounds to the total attenuation length of protons at $z=0$. This plot clearly shows how the photopion production process off the IR/Opt/UV background has little effect on the predicted UHE proton spectra. Indeed, as soon as the pair production threshold with the CMB  is reached, the photopion production with IR/Opt/UV photons is completely subdominant and can be neglected in the  calculation of UHECR with pure proton sources.

The interactions experienced by nuclei with photon backgrounds are different from the proton case. Pair production (for which we use the mass and charge scaling given in \cite{Rachen}) results in a decrease of the Lorentz factor of the UHE nucleus, whereas at higher energies, photodisintegration (also called photoerosion) processes lead to the ejection of one or several nucleons from the nucleus. Different photoerosion processes become dominant in the total interaction cross section at different energies \cite{PSB}.  The lowest energy disintegration process is the Giant Dipole Resonance (GDR) which results in  the emission of one or two nucleons and  $\alpha$ particles. The GDR process is the most relevant as it has the highest cross section with  thresholds  between 10 and 20  MeV for all nuclei. For nuclei with mass $A \geq 9$, we use  the theoretically calculated GDR cross sections presented in \cite{Khan04}, which take into account all the individual reaction channels (n, p, 2n, 2p, np, $\alpha$,...) and are in somewhat better agreement with data than previous treatments.  For nuclei with $A < 9$, we use the phenomenological fits to the data provided by \cite{Rachen}.
Around 30 MeV in the nucleus rest frame and up to the photopion production threshold, the quasi-deuteron (QD) process becomes comparable to the GDR and dominates the total cross section at higher energies.   The photopion production (or baryonic resonances (BR)) of nuclei becomes relevant  above 150 MeV in the nuclei rest frame  (e.g.,  $\sim5\times 10^{21}$ eV in the lab frame for iron nuclei interacting with the CMB), and we use  the parametrisation given in \cite{Rachen} where the cross section in this energy range is proportional to the mass of the nucleus (nuclear shadowing effects are expected to break this scaling above 1 GeV). The reference for this scaling is the deuteron photoabsorption cross section which is known in great detail. It is important to note that photopion cross sections for nuclei are different from the free nucleon case. In particular, in nuclei the baryonic resonances heavier than the first $\Delta$ resonance are far less pronounced than for nucleons and the cross sections are not simply derived from the free nucleon case. We also follow \cite{Rachen} for the treatment of the nucleon multiplicities, the energy losses and absorption probability of the produced pion in the parent nucleus (see below).  Finally, above 1 GeV, the total cross section is dominated by the photofragmentation process which fragments nuclei into individual nucleons or low mass nuclei. For the choices of $E_{max}$ in the present work, the photofragmentation process is negligible.

The contribution of the different photoerosion processes and the different backgrounds to the total mean free path for iron nuclei are displayed in Fig.~\ref{MFP1}b. The photoerosion is dominated by the GDR process  through most of the Lorentz factor range. The baryonic resonances begin to dominate only above $10^{21.5}$ eV where the effect of the GDR starts to decrease. The BR process off the IR/Opt/UV  background does not affect the propagated cosmic ray spectrum, however, as we discuss below,  its contribution to neutrino production is  important. 

\begin{figure*}[ht]\centering\hfill~\hfill\includegraphics[width=6cm]{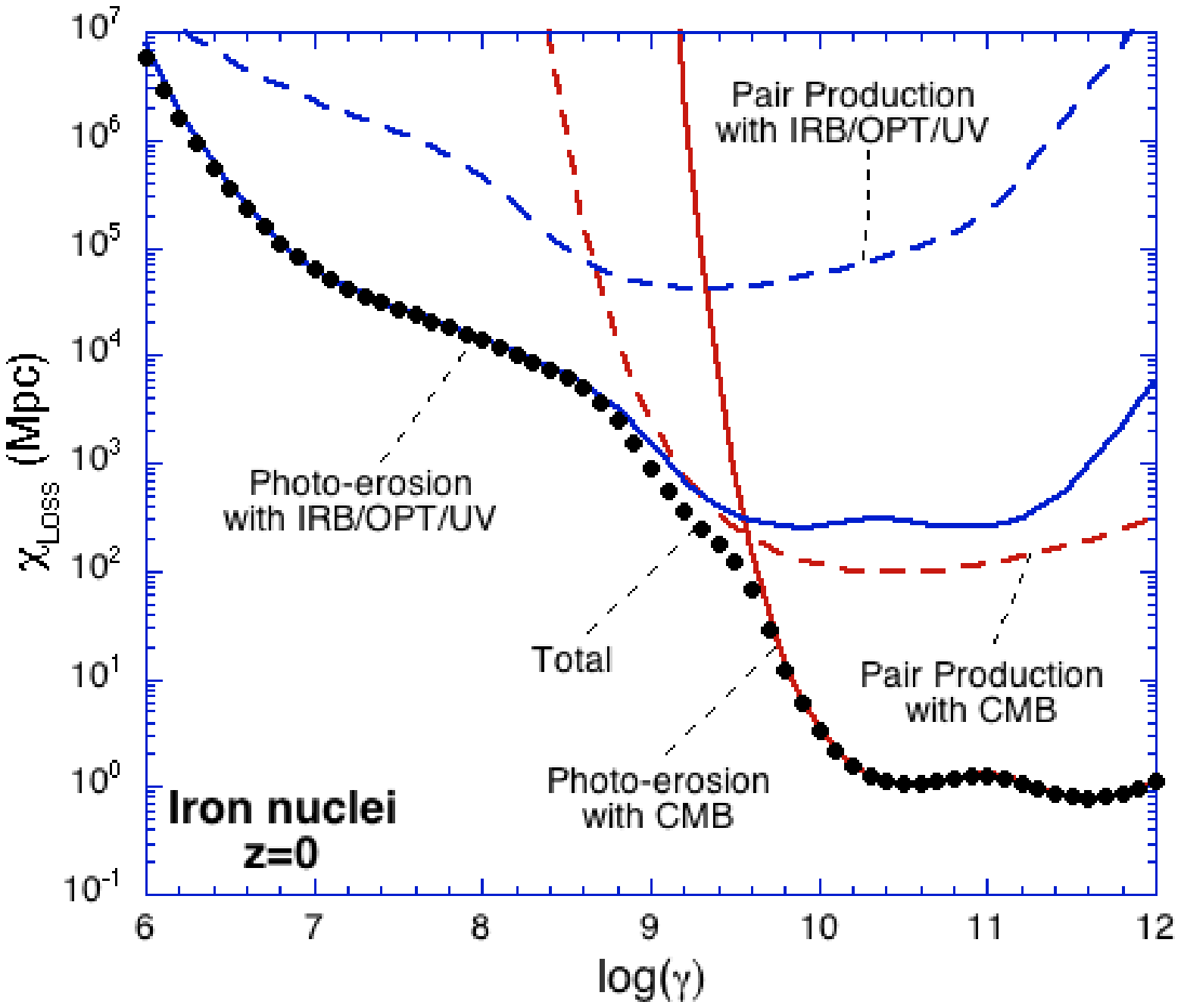}
\hfill\includegraphics[width=6cm]{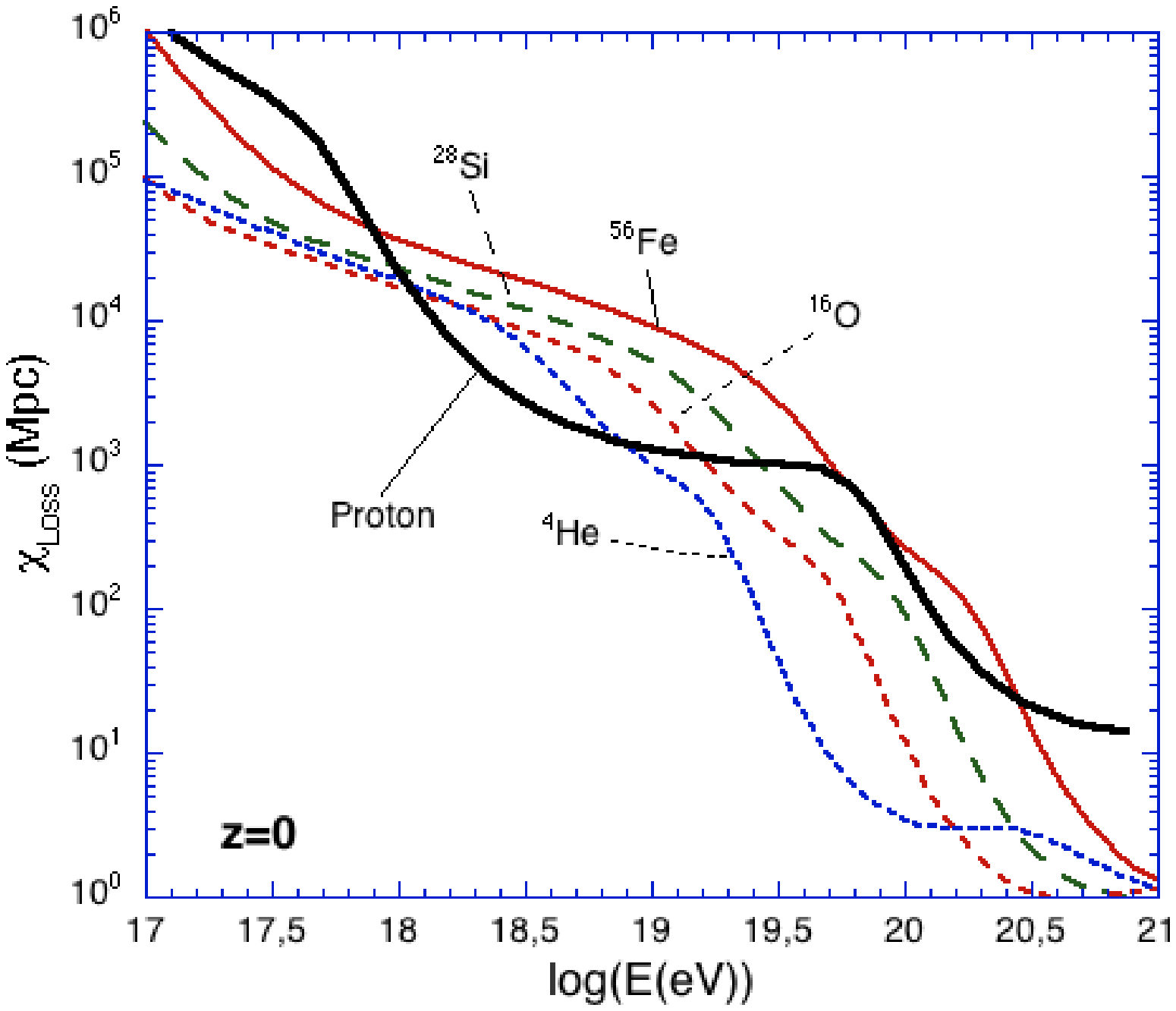}\hfill~\caption{Left: Evolution of the attenuation length of iron as a function of the energy at $z=0$. The contribution of pair production and photoerosion processes off the CMB and IR/Opt/UV photons are separated. Right: Comparison of the attenuation length of different nuclei at $z=0$.}\label{MFP2}\end{figure*}

Fig.~\ref{MFP2}a shows the contribution of pair production and photoerosion processes to the total attenuation length of iron nuclei. Photoerosion processes dominate through most of the energy range and the effect of  pair production is small at low redshifts. Although the competition between pair production off the CMB and photoerosion processes with IR/Opt/UV photons depends on the redshift (e.g., at high redshifts pair production increases due to the strong evolution of the CMB), the propagation of nuclei is mainly dominated by photoerosion processes.  A comparison between the attenuation lengths of different species is displayed in Fig.~\ref{MFP2}b. The figure shows what is
known since \cite{PSB}, that the attenuation length of low mass nuclei are smaller than that of protons and heavy nuclei and, as a consequence, light nuclei should not  contribute as significantly at  the high energy end of the spectrum. Furthermore,  iron nuclei have larger or similar attenuation lengths  to protons up to $3 \times 10^{20}$ eV. However, the energy loss processes are different for protons and nuclei and the sole comparison of attenuation lengths can be misleading. Most of the energy losses of nuclei result in nucleon ejection, thus, unlike protons, a given nucleus does not remain on ``the same attenuation length curve" during its propagation. 
Therefore, in a mixed composition model, the iron component starts to drop around $3-4\times 10^{19}$ eV due to the reduction of the GDR mean free path off far IR photons, whereas the proton component  starts to increase.  
 
We use a Monte-Carlo code to propagate nuclei from the source to Earth as described in detail in  \cite{Allard05a}. The neutrinos produced during the propagation of each particle are
recorded with their flavor, energy, and production redshift. Neutrinos are produced by the decay of pions and of secondary neutrons. In the case of secondary neutron decay,  $n\rightarrow p+e^{-}+\overline{\nu}_{e}$, the energies of the outgoing particles are calculated with a three body decay algorithm.

Neutrinos are also produced by the photopion production of protons and neutrons. This process has been treated in great detail in \cite{EnSecSta}, where all the relevant baryonic resonances and  possible mesons and multi-pions channels were taken into account using \cite{sophia}. In this work, we use a simpler treatment  assuming that the total cross section is dominated by the emission of single pions and use a classical model of the kinematics of the delta resonance process: $N+\gamma\rightarrow\Delta\rightarrow N^{\prime}+\pi$ (where $N$ is a nucleon). For the energy range we consider, the delta resonance dominates multi-pion production and most neutrinos are produced close to threshold \cite{ST68}, \cite{FWS}. As we discuss below, the shape of the neutrino spectrum we obtain closely follows the more detailed treatment of \cite{EnSecSta} (see, e.g., Fig.~\ref{EnSecStaFig} ). 
Photopion production  through the delta resonance has a 1/3 probability  of isospin flip of the incoming nucleon, and each isospin flip leads to the production of three neutrinos. 
For example, in the case of proton interactions producing $\pi^{+}$, $p+\gamma\rightarrow\pi^{+}+n$, the $\pi^{+}$ decay  generates
one $ \nu_{\mu}$, one $\nu_{e}+$ and one $\overline{\nu}_{\mu}$. 
  The charged pion  decay,  $\pi^{+}\rightarrow\mu^{+}+\nu_{\mu}$, 
is  calculated using the two body decay algorithm, while for  the muon decay  
the energies are calculated using the three
body decay algorithm: $\mu^{+}\rightarrow e^{+}+\nu_{e}+\overline{\nu}_{\mu}$.

We use the model above to calculate the neutrino production from interactions of primary and secondary  nucleons and nuclei with the CMB and IR/Opt/UV photons. 
In the case of nuclei propagation,  neutrinos can also be produced via the photopion production of nuclei. The treatment of this component is more uncertain complicated by pion interactions within the nucleus.  To a good approximation,  the pion production can be treated as the production of free nucleons, but the energy losses of the pion inside the nucleus have to be taken into account.  The transfer of the initial kinetic energy of the pion to the spectator nucleons is partly responsible for the high multiplicity of ejected nucleons from this process \cite{Rachen}. Furthermore, the produced pion can also be absorbed by a pair of nucleons before leaving the nucleus. To model these different effects, we follow \cite{Rachen} and use an absorption probability for the produced pion which is proportional to the nucleus radius $P_{abs}=0.13A^{1/3}$. This  approach reproduces well the available data  for a small number of nuclei. If the pion is not absorbed, we remove 40 MeV of kinetic energy in the nucleus rest frame per participant nucleon before calculating the energy of the produced neutrinos. Although this simple treatment allows a reasonable estimate of the  neutrino flux due to photopion production, a more sophisticated treatment would be useful for more precise calculations in the future.

\section{Neutrino fluxes for cosmological distributions of sources}

\subsection{Fit to the observed UHECR spectrum}

We used our monte-carlo simulation to calculate the differential flux of neutrinos for different cosmological distribution of sources.  We propagate
samples of $10^{8}$ particles between $E_{min}=10^{16}$
eV and $E_{max}=Z\times10^{20.5}$ eV. The particles for sources between $z=10^{-5}$
and $z=6$ are followed down to $10^{15}$  eV where  the interaction probability becomes negligible. We inject protons and a  mixed  composition of nuclei and propagate the particles assuming different  source evolutions as described above. 
For each composition hypothesis,  the nominal
value of  the injection spectral index, $\beta$, is chosen by requiring a good fit between the simulated and the observed UHECR spectrum.  
Since  the neutrino production is calculated within the same Monte-Carlo code, the normalization of the simulated spectra to the observed one  provides the normalization for the neutrino fluxes. We normalize the UHECR flux to the HiRes \cite{Bergman+05}  and Auger  \cite{PaulS} spectra, which are a factor of 1.8 below the AGASA \cite{AGASA} spectrum at $10^{19}$ eV. 

\begin{figure*}[ht]\centering\hfill~\hfill\includegraphics[width=6cm]{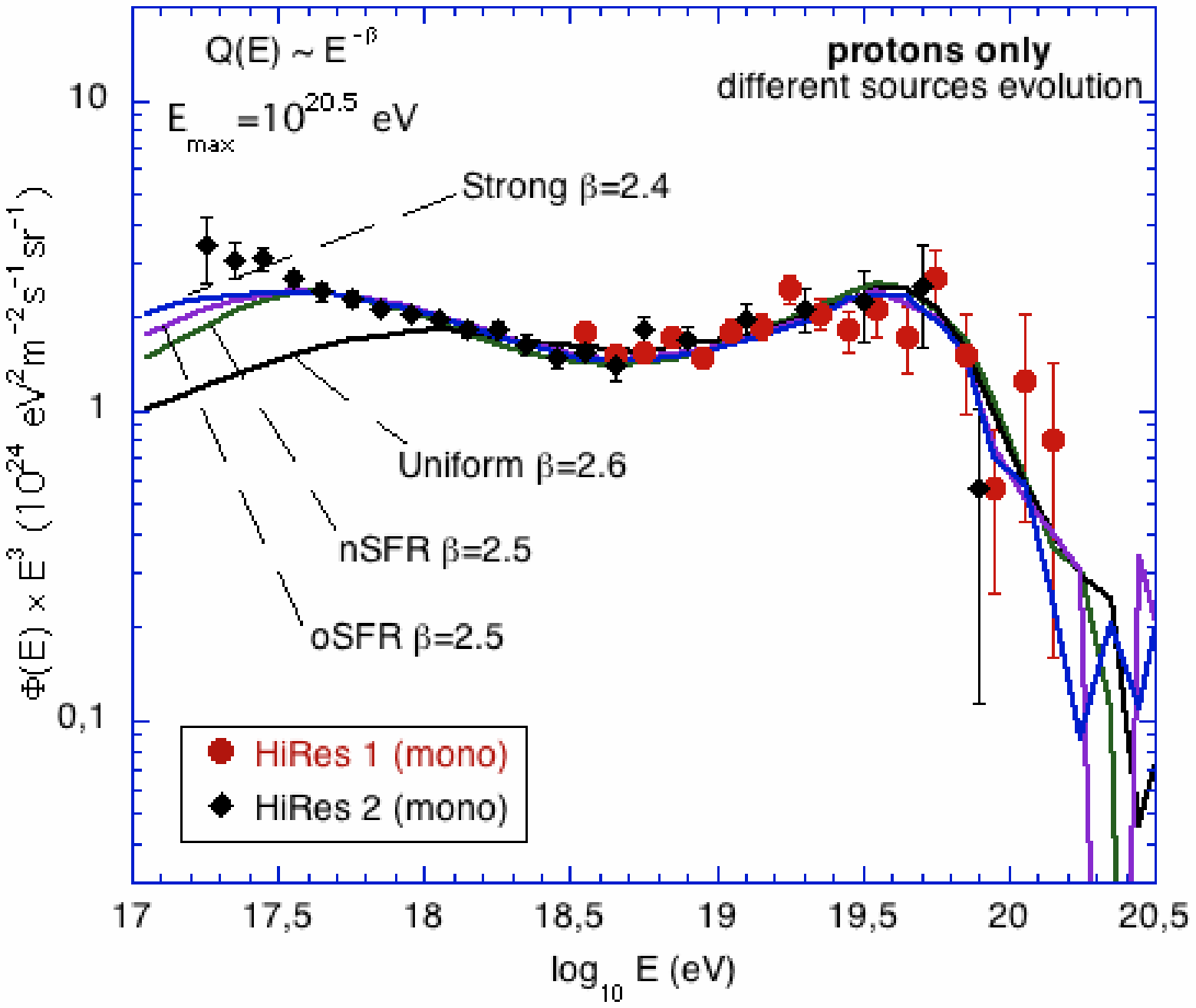}
\hfill\includegraphics[width=6cm]{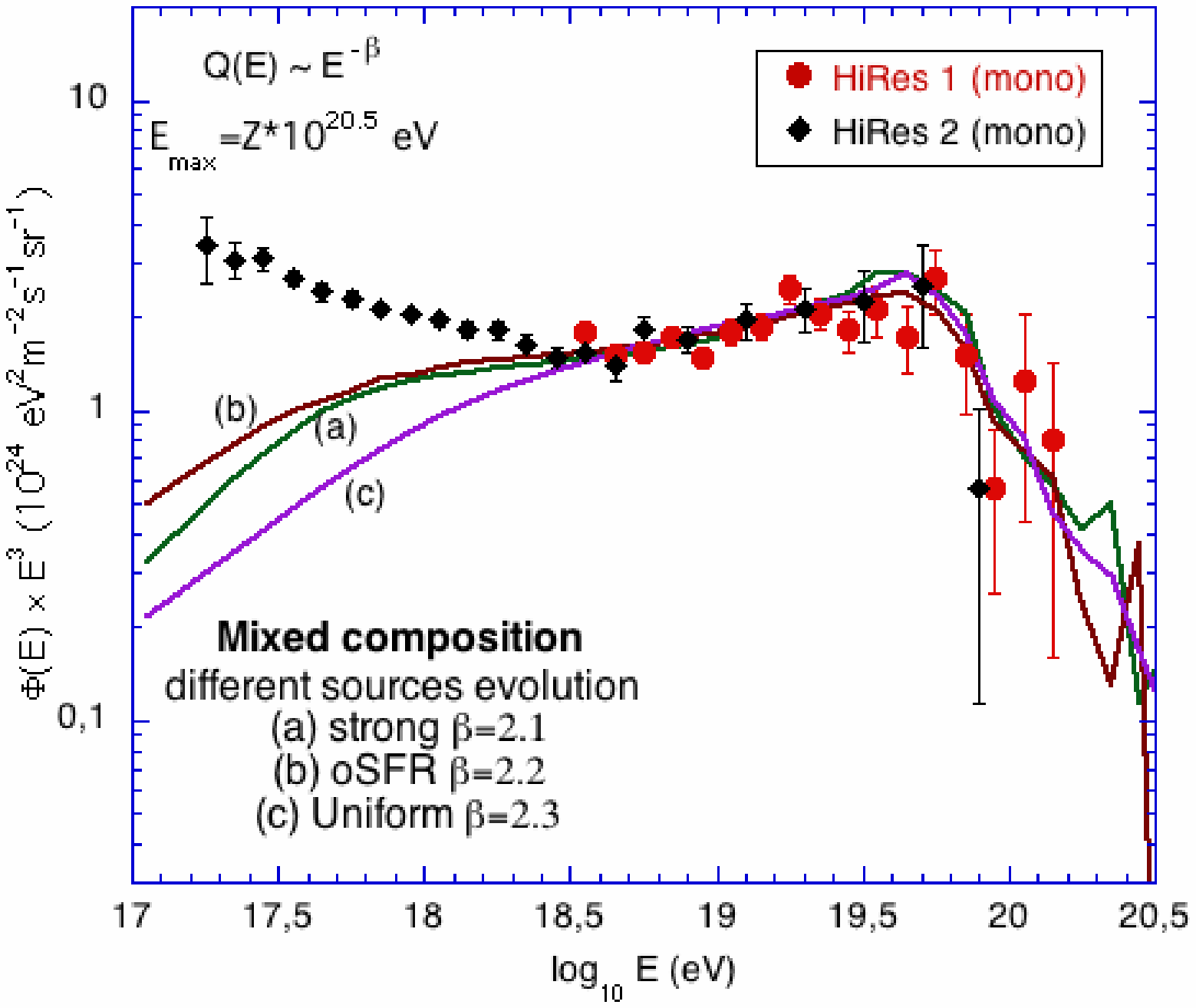}\hfill~\caption{Left: Propagated
spectra obtained assuming a pure proton composition and four  different source evolution hypotheses compared to the HiRes monocular data at ultra high energy. Right: Same as left but assuming a mixed composition and three source evolution hypotheses.}\label{spectra}\end{figure*}

The propagated UHECR spectra that best fit the data for both composition hypotheses and different source evolutions are displayed on Fig.~\ref{spectra}.   
As shown in previous studies (e.g., \cite{bere2,DDMS05}), the generated proton spectra and the presence of a pair production dip are only mildly dependent on the source evolution hypothesis. The amplitude and the energy of the pair production dip minimum (around $10^{18.7}$ eV) only slightly depend on the evolution. However, the beginning of the dip is determined by the transition between the adiabatic and the pair production losses, which is more sensitive to the density of sources at high redshifts \cite{bere2}.   We assume that the galactic component compensates for the difference between the predicted flux and the observed one at energies below the dip. We discuss the implications of the different galactic to extragalactic transition and source evolution models for the neutrino flux below (\S 4.2).

In the mixed composition case \cite{Allard05a}, the transition from galactic to extragalactic components ends at the ankle. In Fig. ~\ref{spectra}b we show the results for three source evolution models (uniform, oSFR, strong). The transition point and the spectrum above $10^{18.5}$ eV is quite insensitive to the source distribution. At energies below the ankle,  the galactic fraction that completes the total observed spectrum depends on the source evolution,  as in the case of pure proton models below $10^{18}$ eV. 
Although the shape of the predicted spectra and their  implications for composition  are not dependent on the source evolution, the spectral indices required to fit the data are harder for stronger source evolution. 
For the  pure proton model $\beta$ moves from 2.6 in the uniform evolution case  to 2.4 for the strong evolution case ($\beta$=2.5 for  oSFR and nSFR) 
\cite{DBO03,bere1,DDMS05,Ahlers05,SS05}, while in the mixed composition  scenario $\beta$ varies from 2.3 for the uniform evolution case  to 2.1 for the strong evolution case ($\beta$=2.2 for  oSFR and nSFR).

\subsection{Neutrino fluxes for pure proton sources}

\begin{figure*}[ht]\centering\hfill~\hfill\includegraphics[width=6cm]{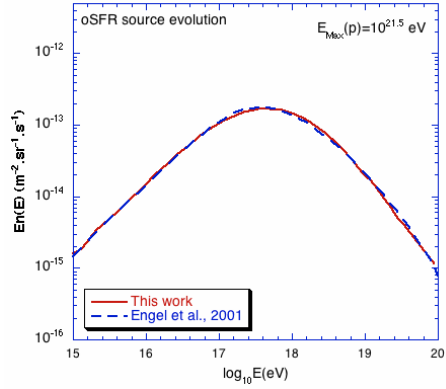}
\hfill~\caption{Comparison of the muon neutrino flux calculated in the present work and in \cite{EnSecSta} for the oSFR source evolution, a spectral index $\beta=2.0$, and $E_{max}=10^{21.5}$ eV.}\label{EnSecStaFig}\end{figure*}

For each composition and source evolution hypothesis our propagation code calculates a normalized flux of neutrinos produced during the propagation. We assume that the produced neutrinos only lose energy due the expansion of the Universe and we neglect their interactions with other particles.
Before estimating the neutrino flux for a given UHECR spectrum, we checked that our simplified treatment of the neutrino production, which uses only the single pion production, gives an accurate estimate of the neutrino flux. We calculated the neutrino flux with our code under the same assumptions of \cite{EnSecSta}, i.e., only interactions with CMB photons, the oSFR source evolution model, a injection spectral index $\beta=2.0$, and a maximum energy at the source of $E_{max}=10^{21.5}$ eV.  A comparison between the two studies is displayed in Fig.~\ref{EnSecStaFig}, where the flux of $\nu_{\mu}$'s produced by photopion interactions is plotted. The differences between the two calculation are very small never exceeding  $\sim 10\%$. This comparison shows that our approximation is accurate enough for a comparison between neutrinos from different injected compositions,  especially considering the other uncertainties in the assumptions (e.g., spectral index, $E_{max}$, source evolution, and magnetic fields).

\begin{figure*}[ht]\centering\hfill~\hfill\includegraphics[width=6cm]{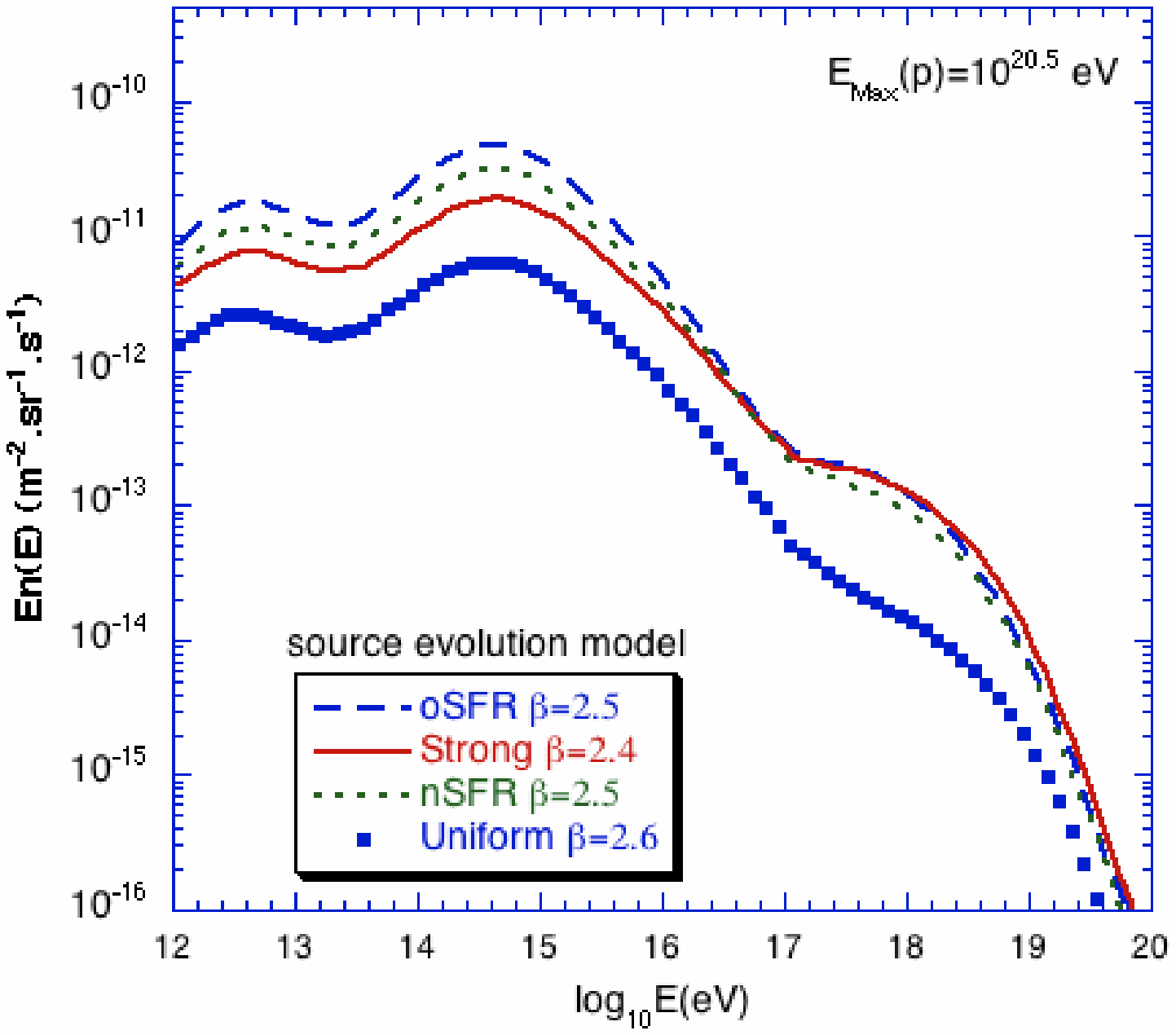}
\hfill\includegraphics[width=6cm]{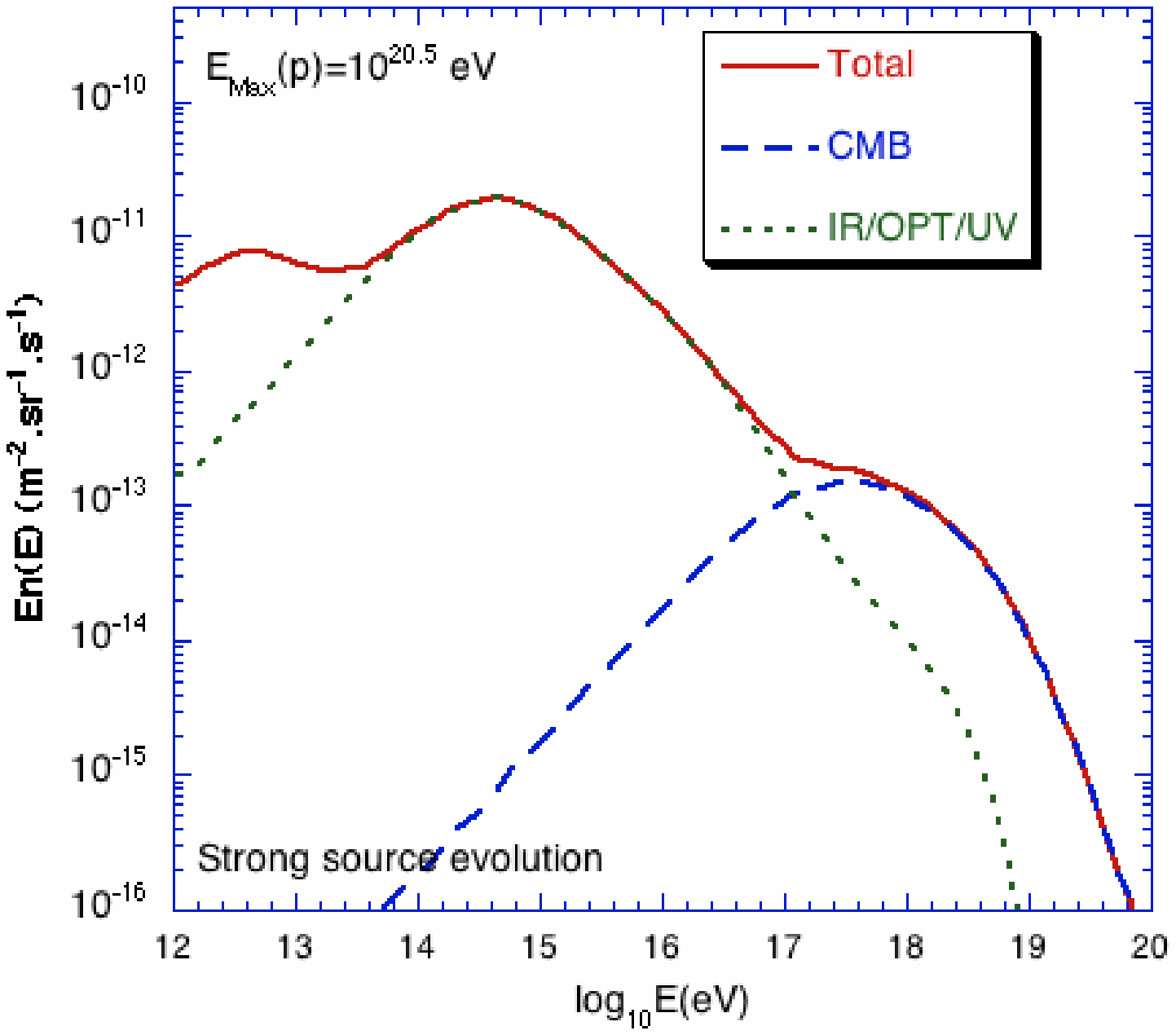}\hfill~\caption{Left: Total neutrino flux from pure proton sources and four source evolution hypotheses.  Right: Contribution of  the different backgrounds to the neutrino flux for a strong source evolution hypothesis. The photopion interactions with CMB (dashed line) and IR/Opt/UV (dotted line) are shown and the difference between the sum of the  two  contributions and the total is the neutron decay component.
 }\label{NeutFluxProt1}\end{figure*}

Neutrino fluxes for a pure proton composition and four different  source evolutions are displayed on Fig.~\ref{NeutFluxProt1}a where it is clear that  the source evolution is a critical parameter.  At high energies,  $\sim 10^{18} $ eV, the neutrino flux for a uniform distribution of sources is almost one order of magnitude below the other hypotheses  \cite{SecSta,bereneut}.
If the sources of UHECRs are distributed uniformly in redshift, the observation of cosmogenic neutrinos will be very challenging  unless the sources can reach extremely high maximum energies \cite{bereneut}. For the three other source evolutions, the neutrino fluxes are comparable: the oSFR source evolution gives a stronger weight to high redshift sources, but in the strong evolution case the harder spectral index counterbalances the redshift effect. 

The contribution of the CMB and IR/Opt/UV backgrounds is detailed for the strong evolution case in Fig.~\ref{NeutFluxProt1}b. At low energies, the neutrino flux is dominated by the contribution of the IR/Opt/UV backgrounds \cite{DDMSS05} and the flux is much higher than at higher energies.  Although the interaction probability with the IR/Opt/UV  backgrounds is much lower than with the CMB, the number of particles that are able to interact is much higher due to the steep injection spectra provided by acceleration mechanisms in astrophysical sources.  The CMB contribution generates a peak at $\sim10^{17.6}$ eV, while the peak at $\sim10^{14.5}$ eV is due to the IR/Opt/UV contribution. The position of the peaks depends on the combination of the evolution of the interaction probability and the injection spectrum.  In Fig.~\ref{background}b, above the threshold, the mean path of protons (thus, the interaction probability) evolves very slowly with energy. As the injection spectra are steep (i.e., more particles are injected at low energies), the number of protons interacting by photopion production with IR/Opt/UV photons is continuously decreasing above the threshold ($\sim10^{16}$ eV at $z=1$), the neutrino peak is then due to interactions close to the threshold corresponding  to neutrino energies around $10^{14.5}$ eV. The interaction probability dramatically increases at higher energies, above the interaction threshold with CMB photons, resulting in a break of the evolution of the flux and a high energy neutrino peak. The peak at high energy  is mainly due to neutrinos produced close to the threshold, therefore, the energy of the peak is only mildly dependent on the maximum energy at the sources. Due to the slow evolution of the IR/Opt/UV background with redshift, the effect of source evolution on the neutrino flux is milder than for the CMB neutrinos, but the influence of the spectral index is higher. The neutrino flux at lower energies are higher for steeper injection spectra, which explains why the oSFR and nSFR evolution scenarios generate higher flux than the strong evolution case, as shown in \ref{NeutFluxProt1}a. 

\begin{figure*}[ht]\centering\hfill~\hfill\includegraphics[width=6cm]{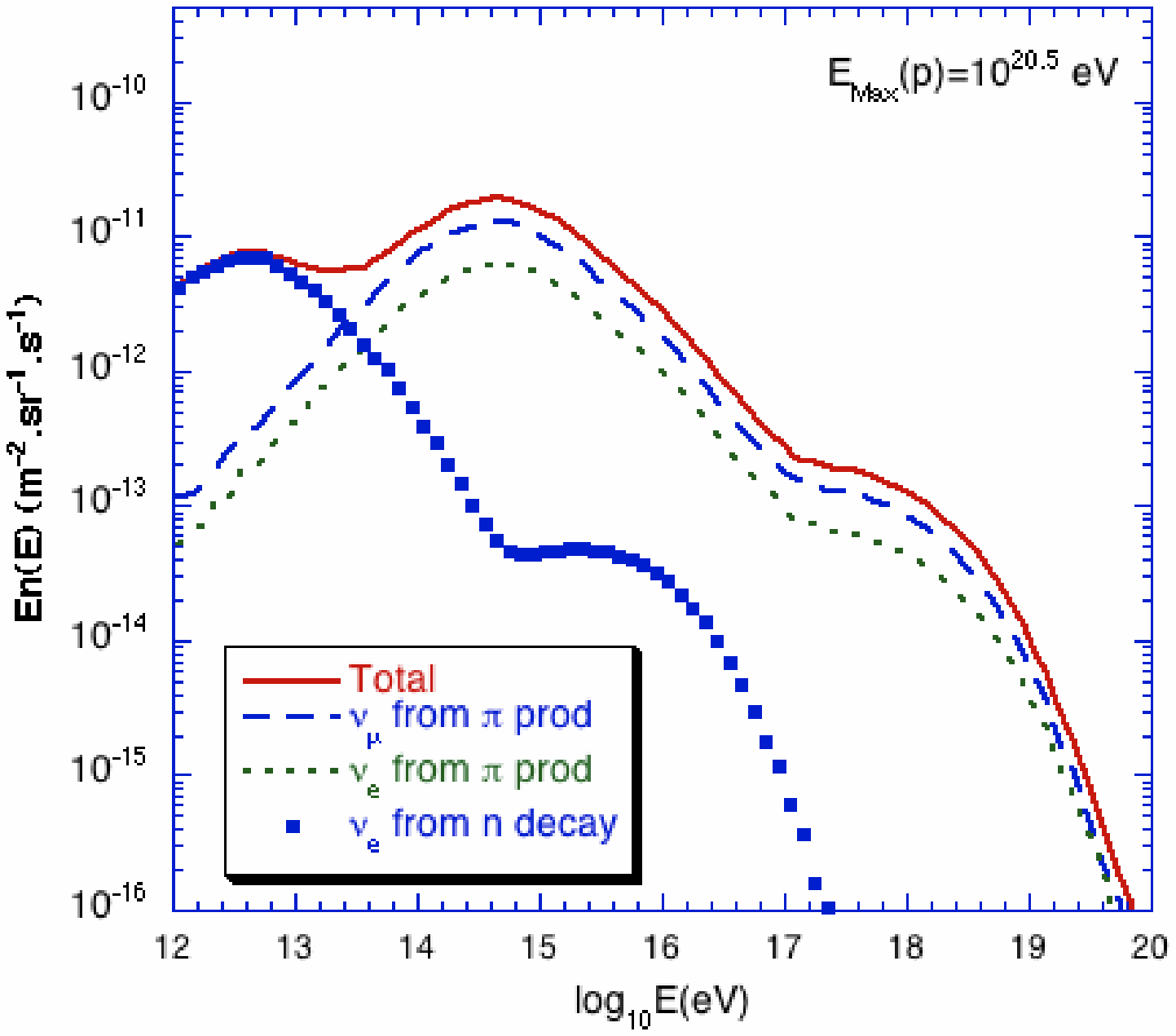}
\hfill\includegraphics[width=6cm]{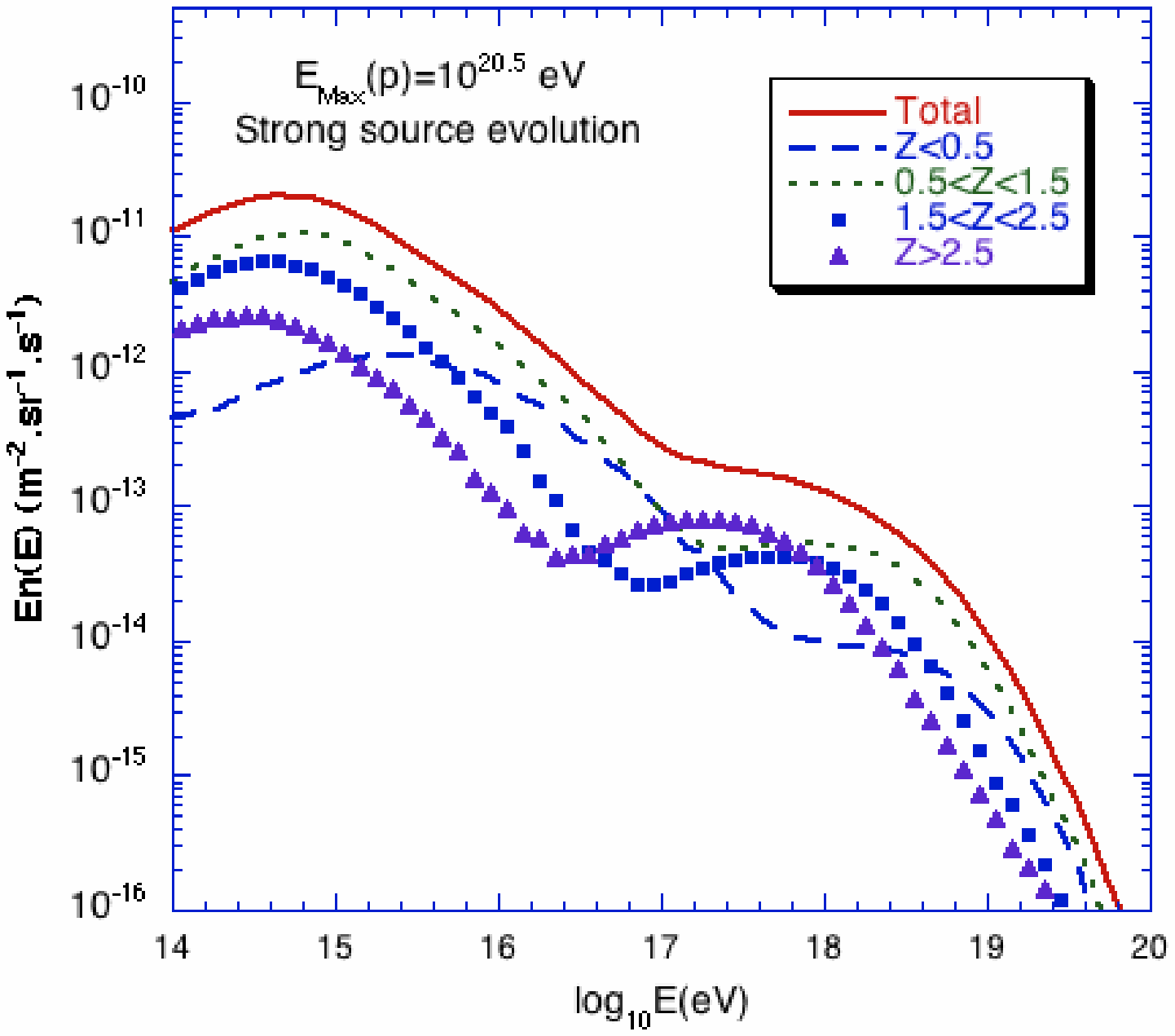}\hfill~\caption{Left: Contribution of different neutrino flavors to the total flux for strong source evolution. Plotted are $\bar{\nu}_{e}$'s from neutron decay and $\nu_{\mu}$ (+ $\bar{\nu}_{\mu}$) and $\nu_{e}$  (+ $\bar{\nu}_{e}$) from pion production.  Right: Contributions from  different redshifts  to the neutrino flux for strong source evolution. }\label{NeutFluxProt2}\end{figure*}

The generated flux of different flavors of neutrinos is displayed on Fig.~\ref{NeutFluxProt2}a for a strong source evolution hypothesis. (Note that the observed fluxes will be affected by mixing.) Each component  ($\nu_{\mu}$ + $\bar{\nu}_{\mu}$ and $\nu_{e}$+ $\bar{\nu}_{e}$ from pion production and $\bar{\nu}_{e}$ from neutron $\beta$-decay)
exhibits a double peak structure due to the contribution from the CMB at high energies and the IR/Opt/UV backgrounds at low energies.  The neutron decay component peaks are shifted to lower energies due to the kinematics of the process. The contribution at low energies coming from neutron decay after proton interactions with UV photons induces the third peak in the total neutrino flux around $10^{12.5}$ eV.

The contribution from  different redshifts to the production of cosmogenic neutrinos is displayed on Fig.~\ref{NeutFluxProt2}b for the strong source evolution hypothesis. It is clear that the contributions from different redshifts depend on the chosen background since the backgrounds evolve differently with redshift. At high energies, the contribution from high redshifts ($z> 1.5$) is important due to the strong evolution of the CMB and the sources. The peaks generated at  different redshifts move in energy mainly due to the evolution of the temperature of the background that allow lower energy protons to interact at higher redshifts.
The strong evolution of the neutrino production with redshift explains the slow increase of the neutrino flux below the high energy peak. In the high energy peak region (around $10^{17.6}$ eV) most of the neutrinos are produced at reshifts above 1.5. In the case of the IR/Opt/UV component, the dominant contribution at the low energy peak comes from intermediate redshifts (between 0.5 and 1.5) as the slow evolution of the backgrounds cannot compensate for  the decrease of the number of sources at high redshifts. For the three redshift ranges above $z=0.5$, the neutrino flux exhibits peaks at similar energies (the differences are due to adiabatic loses rather than the evolution of the background which is mild in the UV range for $z>$1). The low redshift contribution ($z<0.5$) is quite different, as the flux peak from starlight is shifted to higher energies, due to the low-redshift drop in UV emission from young stars evolving off the main sequence.

\subsection{Neutrino fluxes for a mixed composition}

\begin{figure*}[ht]\centering\hfill~\hfill\includegraphics[width=6cm]{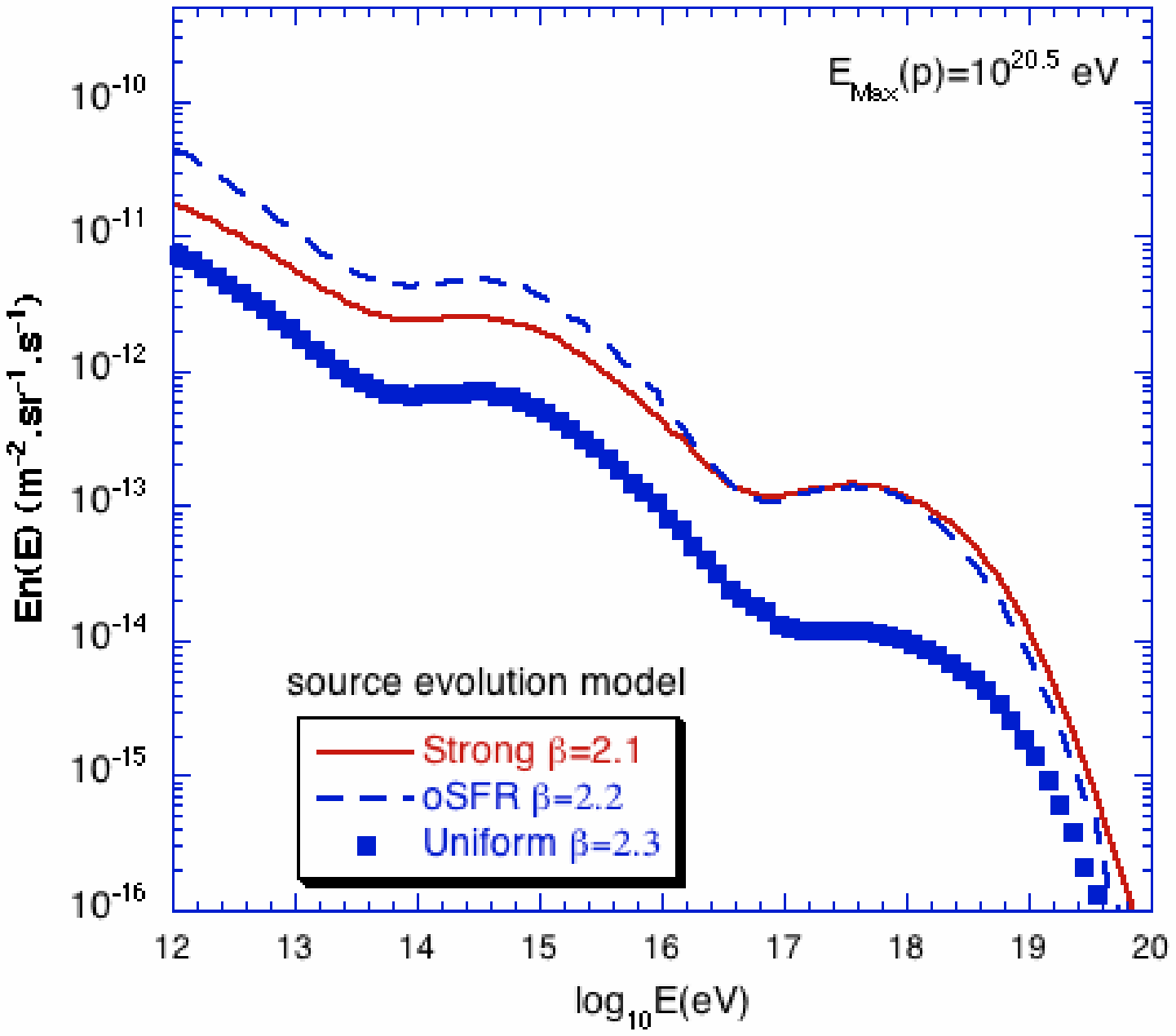}
\hfill\includegraphics[width=6cm]{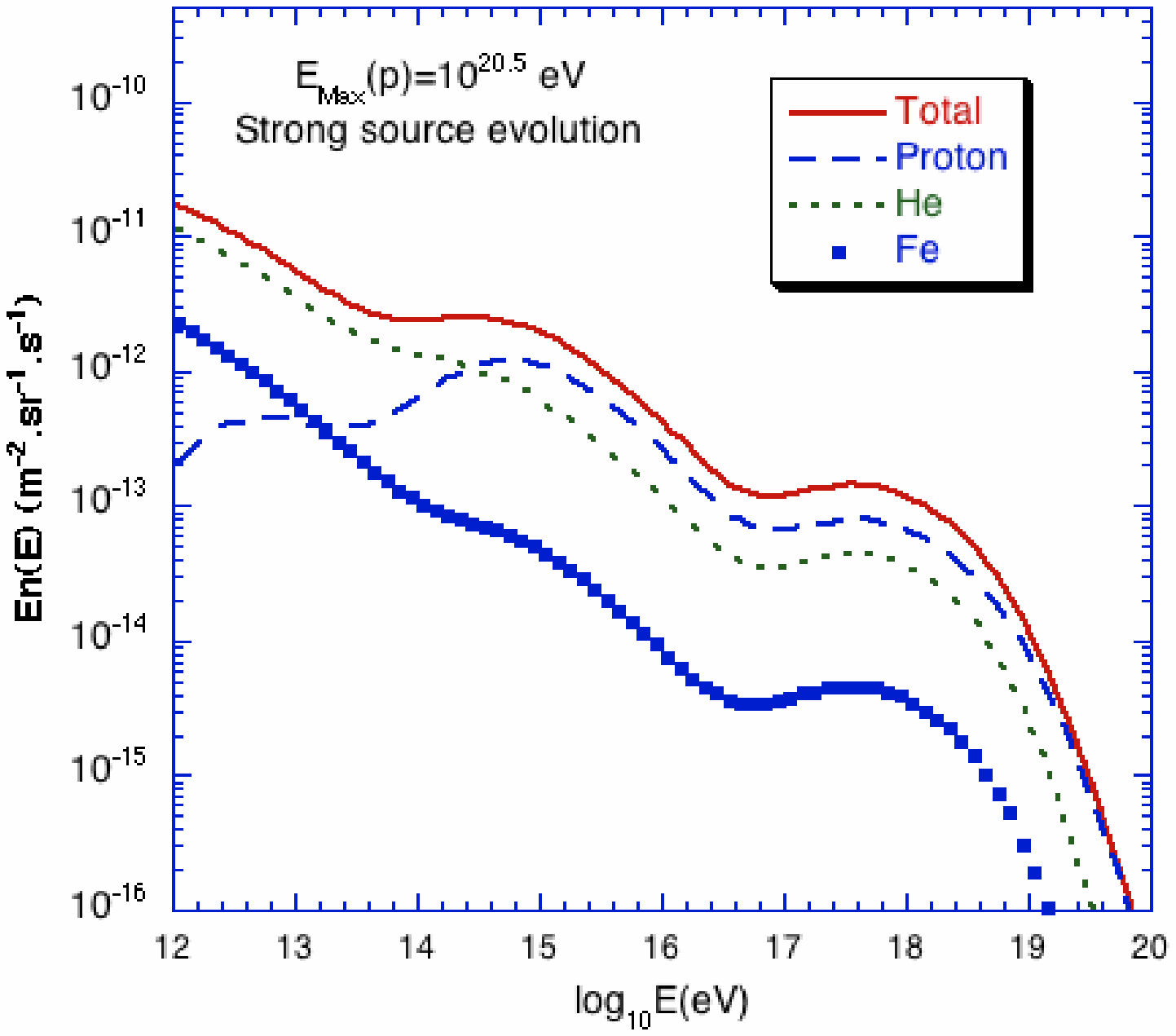}\hfill~\caption{Left: Total neutrino fluxes corresponding a mixed composition and three source evolution hypotheses (the nSFR hypothesis is omitted).  Right: Contribution of
 the different species to the neutrino flux for a strong source evolution hypothesis 
 }\label{NeutFluxMix1}\end{figure*}

In the case of a mixed composition, the expected neutrino flux is shown in Fig.~\ref{NeutFluxMix1}a for the different source evolution hypotheses. At high energies, the flux is very similar to the pure proton case, which is not surprising since the mixed composition models with $\beta$ between 2.1 and 2.3 are proton dominated (see Fig.~\ref{Abond}). The flux is much higher in the strong evolution case  compared to a uniform distribution.  There is very little difference between the strong and oSTR evolution cases (for simplicity, we omit the nSFR). The intermediate energy peak appears around $10^{14.5}$ eV and the neutron decay peak is just below $10^{12}$ eV.

In the case of a strong evolution, the contribution of protons, He, and Fe nuclei to the total flux are displayed on Fig.~\ref{NeutFluxMix1}b. Above $10^{15}$ eV, the main contribution is due to protons ($\sim 65\%$), the second contribution comes from He,  and Fe nuclei contribute only of a few percent over the whole energy range. 
Due to the harder spectral index, the intermediate energies peak for the proton component in Fig.~\ref{NeutFluxMix1}b is slightly shifted towards higher energies (to $\sim 10^{15}$ eV), when compared to the corresponding peak  in the pure  proton case. The intermediate energy peak for the total flux from all the species is again around $\sim 10^{14.5}$ eV due to the contribution of the other species which become dominant for energies below the proton peak. 
  The neutron decay peak of the proton component is invisible on the total flux as it is completely dominated by the nuclei contribution in this energy range. It is important to note however that the contribution of the different species is strongly dependent on the relative abundances assumed at the source. In our model, protons contribute $\sim50\%$, helium $\sim30\%$, and  Fe $\sim5\%$ at the source for a spectral index $\beta=2.1$.  
 
\begin{figure*}[ht]\centering\hfill~\hfill\includegraphics[width=6cm]{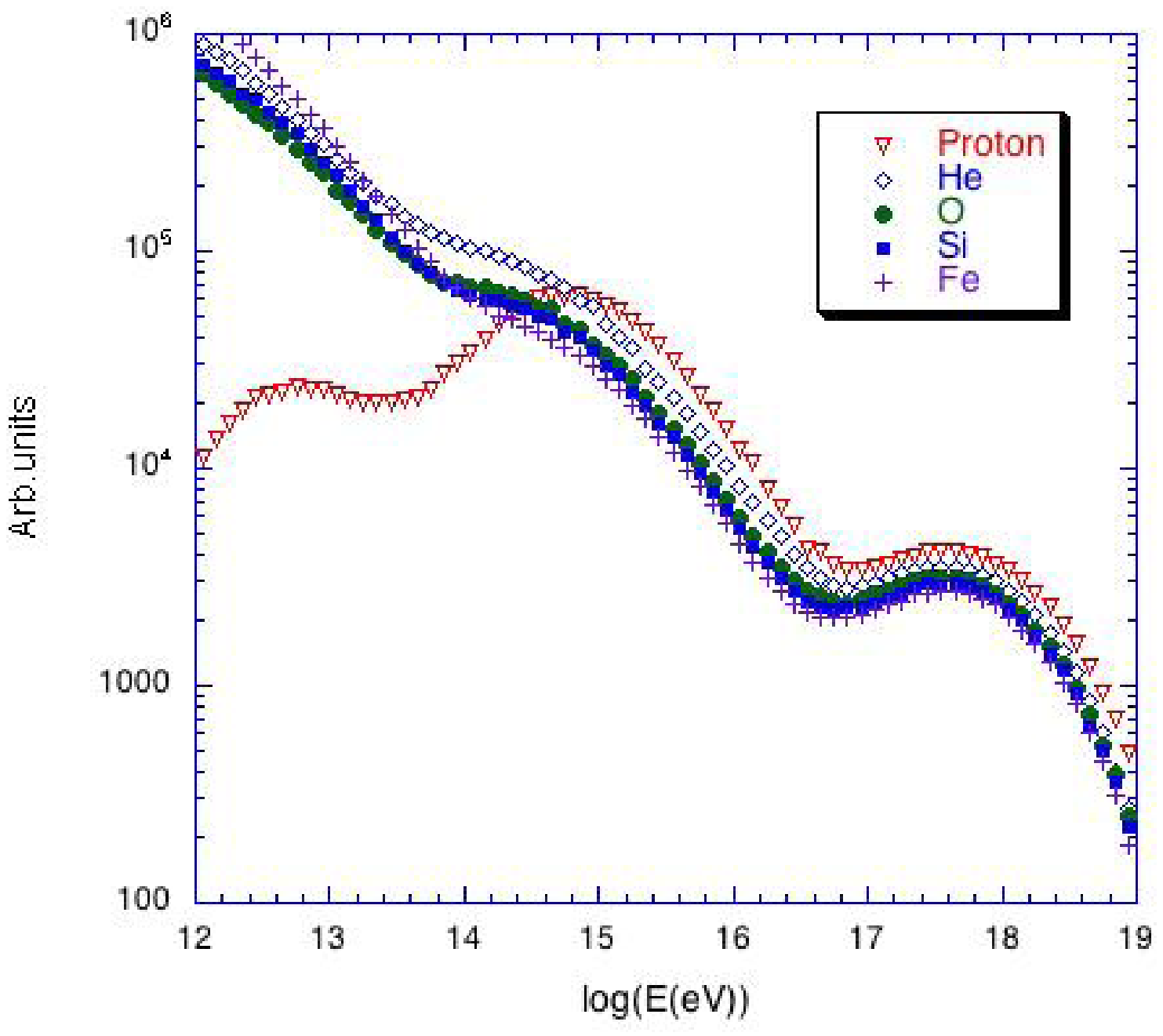}
\hfill\includegraphics[width=6cm]{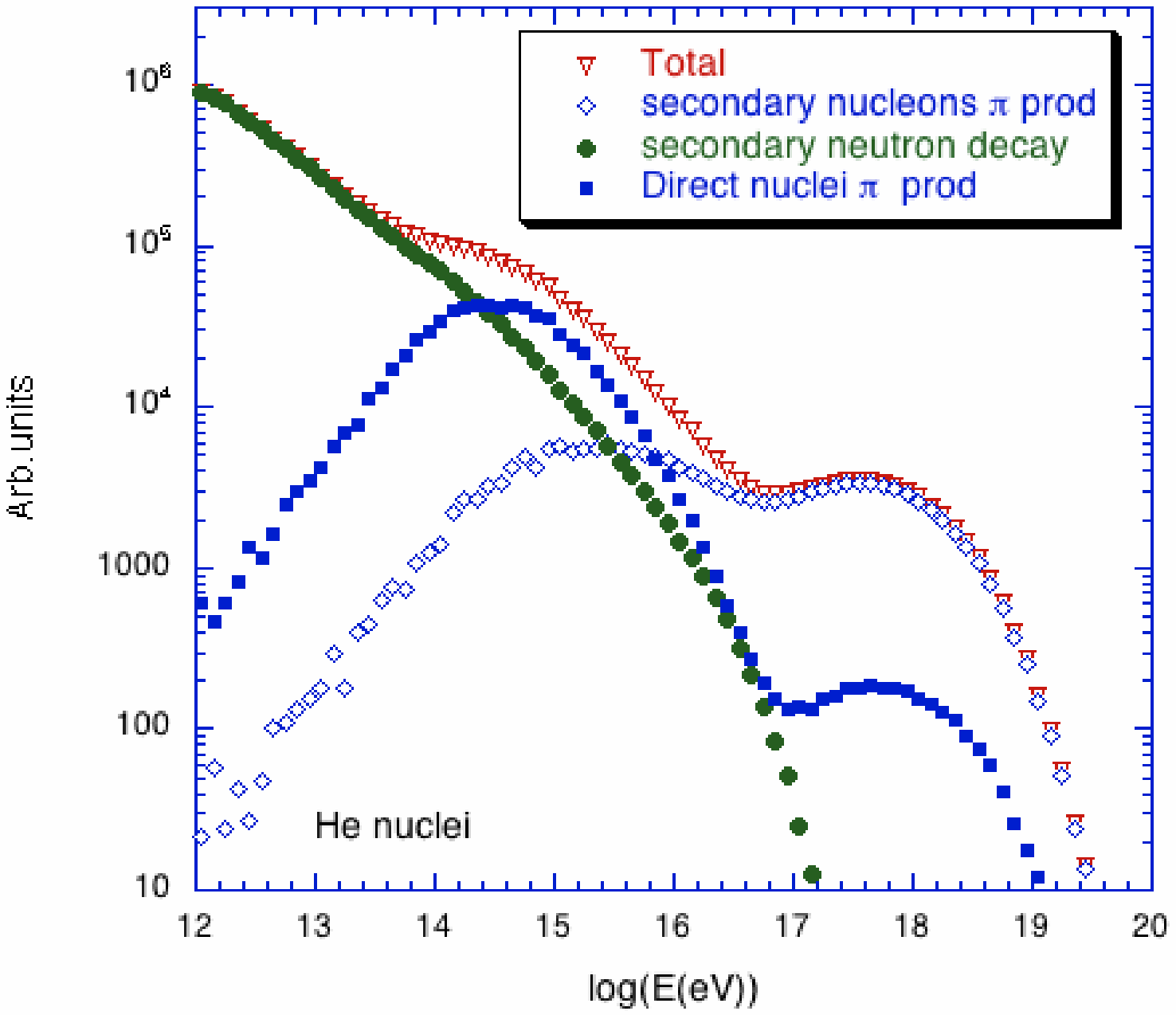}\hfill~\caption{Left: Comparison of the neutrino production of different species assuming the same integrated abundances for all species between $10^{16}$ eV and $Z\,10^{20.5}$ eV, a strong source evolution scenario, and  $\beta= 2.1$.  Right: Contribution of the different processes of neutrino production for the He generated neutrino flux in the left panel. }\label{NeutFluxMix2}\end{figure*}

To show the relative contributions to the neutrino flux from different species, we show in Fig.~\ref{NeutFluxMix2}a the generated fluxes in arbitrary units  assuming the same spectral index and similar integrated abundances at the sources between $10^{16}$ and $Z \, 10^{20.5}$ eV for all the species. 
It is clear that, except for protons,  there is very little difference between the contributions from different species under the assumption of a maximum energy scaling with $Z$. 
As mentioned in \cite{Allard05b}, the acceleration of nuclei above $10^{20}$ eV is not required to fit the UHECR spectrum if protons are accelerated above $\sim10^{20}$ eV, therefore, the contribution of nuclei at high energies is uncertain and strongly dependent on the maximum energy reached at the source. 

In Fig.~\ref{NeutFluxMix2}b we show the contribution of the different production processes for He nuclei which is qualitatively similar to those of heavier nuclei. The high energy neutrino production for nuclei is dominated by the pion production of secondary nucleons off CMB photons, which requires nuclei to be accelerated above $\sim A \times 5 \times 10^{19} \, {\rm eV} /(1+z)$  corresponding to  $E_{\max} \sim 10^{21}$ eV for Fe nuclei. 
For the maximum energy we assumed (see \S 2), the flux from direct photopion production of nuclei is  low due to the fact that this process only becomes dominant above $\sim A\times1.5 \times 10^{20}$ eV and that the produced pion can be absorbed before producing neutrinos (the relative contribution from this process for iron is even lower than in the case of helium). If nuclei are not accelerated to the highest energies the flux is lower than our estimate, while the flux could be slightly higher if the proton abundance is higher than in our model.

In the intermediate peak region around $10^{14.5}$ eV,  the neutrino flux from nuclei originate mainly from direct pion production and neutron decay from secondary nuclei. Direct photopion production is the dominant process in the peak region, which may appear surprising since the GDR interaction probability (which is responsible of the nucleon emission) is  higher than the photopion probability. However, for the photopion production of secondary nucleons, the neutrino emission requires first the emission of a nucleon and a subsequent photopion interaction of the emitted nucleon. At low energies, the realization of both of these requirements becomes less probable than a direct photopion interaction (despite the pion absorption probability), therefore, direct photopion dominates.
The neutrino flux produced by secondary nucleon photopion production drops faster at low energies than in the free nucleon case (see the proton component on  Fig.~\ref{NeutFluxMix2}a).

At the lower energies, below $10^{14}$ eV, the neutrino flux from nuclei is dominated by the contribution of secondary neutron $\beta$-decays. This component is far more important than in the case of primary protons as the interaction probability of nuclei (via the GDR process) is much higher. Unlike the secondary nucleon photopion component, the flux keeps increasing at lower energies as a subsequent interaction of the ejected nucleon is not necessary. The  CMB contribution peaks around $10^{14}$ eV but is overwhelmed by the IR/Opt/UV background contribution in the case of He nuclei. The shift of the neutron decay peak to lower energies compared with the pure proton case can be explained by the fact that the energy per nucleon threshold of the GDR process is lower than for the photopion process. Neutrons with energies below $10^{16}$ eV can be ejected in contrast with the  pure proton case where neutrons are produced following a photopion interaction of a proton, which does not occurs below $10^{16}$ eV even at high redshifts.

Finally,  when comparing the contribution of the different species it is important to note the sensitivity of the generated flux to the spectral index assumed. Indeed, as the threshold energies for the different processes scale approximately with the mass of the nuclei (i.e., they occur at approximately the same Lorentz factor for all species), lower mass nuclei have lower energy thresholds. On the other hand, the cross section and the number of nucleons that can be ejected are higher for higher mass nuclei. 
The  steep spectral indices in astrophysical accelerators together with lower energy thresholds  favor the neutrino production of low mass nuclei, which is why the neutrino production is higher for He nuclei. The evolution of the cross sections and of the number of nucleons with the mass tend to counterbalance partially this effect especially if the spectral index is hard. For $\beta=2.1$ the 
difference between the neutrino production from different  nuclei is relatively small, while for softer spectral indices the lower mass nuclei become even more dominant over the whole energy range.

\subsection{Comparison between the two composition hypotheses}

\begin{figure*}[ht]\centering\hfill~\hfill\includegraphics[width=10cm]{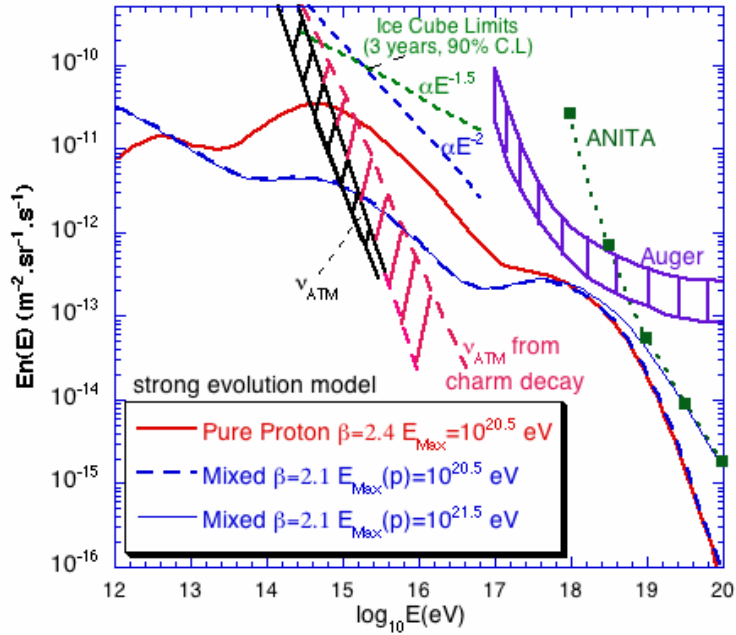}
\hfill~\caption{Predicted cosmogenic neutrino flux for pure proton and mixed composition in the case of the strong source evolution compared with the sensitivities of Auger ($\nu_{\tau}$) and ANITA at high energy  and the limit of Ice Cube (for the $\nu_{\mu}$ detection channel only) after three years of observation with 90 C.L. (assuming neutrino spectra $\propto E^{-2}$ and $E^{-1.5}$) at low energy, estimates of the atmospheric neutrino flux and atmospheric neutrinos due to charmed interactions are also displayed.}\label{NeutFluxFinal}\end{figure*}

The comparison between the pure proton and the mixed composition is displayed on Fig.~\ref{NeutFluxFinal} for a strong source evolution hypothesis.
For the other choices of source evolution, the  comparison is qualitatively similar. At high energies, where detection is feasible, the fluxes are comparable. 
Changes in the detailed composition at the source or in the maximum energy reachable for the nuclei can slightly increase or decrease the predictions, but  for a proton dominated mixed composition the flux at the highest energies is very close to the pure proton model. 
The similarity between the fluxes arises from the fact that the lower contribution of nuclei to the neutrino flux is compensated by the harder spectral index required to fit the UHECR data. 
Conversely, at lower energies, the higher fluxes expected for the pure proton scenario are mainly due to the softer spectral index that give a higher luminosity at low energies. In the intermediate  energy range, detectability is  limited by the atmosphere neutrino background for  energies below  $10^{15}$ eV. For energies between $10^{15}$ eV and  $10^{17}$ eV, the pure proton case increases the chances of future detectability.
However, the comparison between the two compositions is sensitive to a few assumptions as we discuss below.

\section{Discussion}

In the previous sections, we calculated neutrino fluxes for mixed and pure proton compositions for different source evolution hypotheses. In this work,  we made some assumptions prior to the calculation that we further discuss below as they raise some uncertainties on the expected cosmogenic neutrino fluxes. We also discuss the IR/Opt/UV background and the possible detection of the predicted cosmogenic neutrino fluxes.

\subsection{Change of the proton injection spectral index  at low energy}

In the pure proton case, we assumed that the injection spectrum was constant down to low energies. This assumption affects the transition between the galactic and extragalactic cosmic ray components. In particular, the extragalactic proton flux obtained around $10^{17}$ eV is too high to match the low proton abundance observed by KASCADE \cite{Kascade}. A change of the injection spectral index around $10^{18}$, making it harder at low energies, can alleviate the problem with the Kascade data and prevent the sources from being too luminous as proposed by \cite{bere1}. In this model,  the transition between heavy galactic and proton extragalactic components is extremely steep as discussed in \cite{Allard05b}.
The consequence of such a change in the spectral index for the neutrino flux in  the strong evolution case is to bring the pure proton case down to mixed composition case, i.e., a flux well below the one displayed in Fig.~\ref{NeutFluxProt1}.

\subsection{Neutrino fluxes and extragalactic magnetic fields}
Another way to solve the problem of the proton abundance at $10^{17}$ eV is to invoke a magnetic horizon effect \cite{L04,alo} that prevents lower energy extragalactic protons from reaching the Earth.
 The effect on the steepness of the galactic to extragalactic transition is similar to  the proposal discussed above (\S 5.1),  but without the need to change the spectral index.  The neutrino fluxes we calculated would not be significantly changed as the magnetic fields would only have an effect on the charged particles. 

Strong extragalactic magnetic fields can change the propagated UHECR spectrum and may challenge the fits in \S 4, specially for mixed composition models \cite{SiArmen05}. The fairly weak observational and theoretical constraints on the intensity and configuration of extragalactic magnetic fields  make a precise estimate of their possible effects very difficult (see, e.g.,   \cite{dolag04,sigl}). The effect of magnetic fields should be most significant at lower energies, where 
we  assumed that the particles could escape the source for both composition hypotheses down to $10^{16}$ eV. This  assumption is somewhat extreme even for pure protons. Astrophysical sources are expected to be surrounded by magnetic fields and radiation fields that can be quite strong, for example in galaxy clusters.  Charged particles can be confined and the diffuse neutrino flux in the low energy range could be  significantly higher than the fluxes we calculated above.

\subsection{Infrared, Optical, and Ultraviolet Backgrounds} 
In order to calculate intergalactic IR photon fluxes and densities and their evolution over time (or redshift), 
the authors of Ref. \cite{MS05} adopted the approach of using an empirically 
based method of calculating the infrared background radiation based on (1) the
luminosity dependent galaxy spectral energy distributions obtained from 
observations, (2) observationally derived galaxy luminosity distribution 
functions and  (3)   the  latest  redshift-dependent  luminosity
evolution  functions, primarily based on recent data from the {\it Spitzer}
infrared space telescope. As is shown in Figure 6 of that paper, the predicted infrared flux distribution as a function of energy hugs the data from the 
{\it Cosmic Background Explorer} within 1 $\sigma$, the lower limits from galaxy counts, and the upper limit from TeV $\gamma$-ray constraints. The estimated photon 
backgrounds emitted by stars and warmed dust grains in galaxies are now well measured from the current epoch back to a redshift of 1. The uncertaintly in the far infrared 
energy range is $\sim$20\% and that in the mid-IR range is $\sim$30\%. The two different models in Ref. \cite{MS05} differ in the optical-UV range by $\sim$ 50\%. 
These are a fair measures of the 
observational and calculational uncertainties in these energy ranges.  However, at redshifts greater than 1 the uncertainty in the integrated galactic photon emission 
could be significantly larger, as is the uncertainty in the ratio of far-infrared to UV emission from dust absorption and reradiation at these redshifts.
However, the strongest feature in the background photon spectrum, {\it viz.}, the sharp cutoff below the Lyman 
limit responsible for the intermediate peak, has a more solid theoretical and observational foundation.  
None of the dozen or more galaxies imaged above this limit of 13.6 eV energy has any detected continuum radiation, even at the level of a few percent of 
the emission at energies below this cutoff (\cite{MKW}).
Although, the uncertainties on the backgrounds are higher at high redshifts, the evolution of the IR-UV backgrounds with redshift remains milder than in the case of 
the CMB. Therefore the contribution of high redshifts to the neutrino fluxes would anyway not be dominant even with a significantly higher background. As the IR-UV 
backgrounds are well constrained at low and intermediate redshifts ($z \leq 1.5$), we can conclude that the knowledge these backgrounds is good enough not to be a major 
source of uncertainties for our calculations.

\subsection{Detection of the cosmogenic neutrino fluxes} 

The neutrino fluxes that we obtained assuming a strong source evolution for pure proton and proton dominated mixed composition models are displayed on Fig.~\ref{NeutFluxFinal}. In this figure we use the AGASA normalization which gives neutrino fluxes about 80\% higher than our earlier estimates. This 80\% difference between HiRes and AGASA normalization gives a fair estimate of the uncertainties triggered by the choice of the normalization. In the figure we also show the sensitivity of Ice Cube \cite{ICECUBE} for three years of observations and two different assumptions for the spectral index of the neutrino flux: $\beta=2.0$ and $\beta$=1.5, the sensitivity of ANITA \cite{ANITA} for a 45 days flight time, and the sensitivity of Auger to  $\nu_{\tau}$'s \cite{Auger_tau} assuming maximum mixing and a sensitivity ranging from one event a year to one a decade.

For the strong source evolution hypothesis, the flux that we calculated is close to the ANITA sensitivity and would be detectable if the maximum energy of acceleration is high enough, as it can be seen in the case of the mixed composition for $E_{max}=10^{21.5}$ eV. The typical case treated in most neutrino studies assumes a pure proton composition with a spectral index $\beta=2.0$, the oSFR source evolution, and a high value of the maximum energy (typically $10^{21.5}$ eV), which generally give higher fluxes than the ones we calculated here and slightly above the ANITA sensitivity. However, this typical scenario with a hard injection spectral index does not provide the best fit to the UHECR spectrum, especially below $10^{19}$ eV. It thus requires a Galactic component extending up to high energies, with typically equal flux of the galactic and extragalactic components at the ankle. The differences in the neutrino fluxes obtained in each case appear too weak to allow one constraining the nature of the ankle with the sole observation of high energy neutrinos.

Since the predicted fluxes in the case of a mild source evolution are unfortunately too low to be observed by present detectors for reasonable assumptions on the maximum energies, the observation of any UHE neutrinos by Auger or ANITA would put very severe constrains on the source evolution \cite{SecSta}, which would then have to be strong or very strong (and coupled to a high value of the maximum energy). Note that the distribution of plausible sites of UHECR sources such as active galactic nuclei, gamma-ray bursts, and galaxy clusters, evolve strongly with redshift, therefore one can conclude that, despite the high value of maximum energy required, the possibility of detecting neutrinos at high energy is not unlikely.

In the low energy range, the situation is more complicated since the amplitude of the cosmogenic diffuse flux depends on the confinement efficiency at the source. However, the fluxes predicted assuming that the particles propagate freely in the intergalactic space are reasonably high. In Fig.~\ref{NeutFluxFinal}, we compare the predicted fluxes with the 90 $\%$ confidence level limits that Ice Cube \cite{ICECUBE} should provide after three years of observations, for two different assumptions on the spectral index of the neutrino flux, $\beta=2.0$ and $\beta$=1.5, which correspond roughly to the proton and the mixed cases, respectively. Note that the limits shown only consider the $\nu_{\mu}$ channel, whereas Ice Cube should also be sensitive to electromagnetic showers from $\nu_{\tau}$ and $\nu_{e}$ \cite{Kow}. 
Although the fluxes at lower energies are not as close to the instruments sensitivity as they are at high energies, and even more so for a mixed composition, they could be greatly enhanced if particles would be confined for a sufficiently long time around the sources. This is actually easier to achieve in the case of nuclei. Therefore, if Ice Cube observes a diffuse flux, it may be due to the confinement of nuclei in VHE  and UHECR sources.

Due to its large sensitivity, Ice Cube may  be able to constrain the acceleration and the confinement of cosmic rays in sources even if cosmic ray nuclei are present. At low energies, the detection of a diffuse flux is complicated by the presence of the atmospheric neutrino flux and the uncertain flux from charm decay (see Fig.~\ref{NeutFluxFinal}).  Methods to deconvolve atmospheric neutrinos from astrophysical diffuse flux were proposed in \cite{Kow} and may be effective in pulling a diffuse flux in different detection channels. Neutrino point source detections would be even more illuminating allowing for a significant improvement in studying UHECR confinement and sources environment.

The Pierre Auger Observatory \cite{cronin} should be able to study the spectrum and composition of the ankle, a possible GZK feature in the UHECR spectrum  and identify point sources of UHECR. 
The combination of any cosmogenic  neutrino detection with detailed UHECR spectrum, composition  and anisotropy analysis from the Auger Observatory would greatly help solve the mystery of the origin of UHECRs. 
 
\section{Conclusion}
We studied the cosmogenic neutrinos production using a mixed composition hypothesis and several source evolution hypotheses. At high energies, the cosmogenic neutrino flux for our mixed composition model is very close to the pure proton case. In this range,   ANITA and Auger can detect or  strongly constrain the source distribution evolution with redshift \cite{SecSta} independent of the composition model assumed. In the lower energy range, the flux are higher  due to interaction with the IR/Opt/UV backgrounds \cite{DDMSS05}. The effect is strongest for pure protons, due to steeper spectral index, but it is also present in the mixed composition case  mainly due to the direct photopion production process. If the confinement at UHECR sources is efficient, stronger fluxes can be expected in the Ice Cube energy range.
In sum,  Ice Cube, Auger, and ANITA will either detect cosmogenic neutrinos or provide important  constraints on the confinement and the environment of the sources of extragalactic cosmic rays once combined with UHECR composition measurements.  The combination of UHE neutrino and cosmic ray detectors will  open a new window into the highest energy phenomena of the present universe.

\subsubsection*{Acknoledgements:}
We wish to thank Daniel de Marco, Todor Stanev and Ralph Engel for helpful discussions. This work was supported in part by NSF PHY-0457069 and by  the KICP under NSF PHY-0114422
at the University of Chicago.

\end{document}